\documentclass[11pt]{article}
\usepackage{color}
\usepackage{amsmath,amsfonts,amssymb,stackengine,graphicx,epsfig,latexsym}
\usepackage{fullpage}
\usepackage[T1]{fontenc}
\usepackage{hyperref}
\usepackage{amssymb,amsmath}

\newcommand{\cA}{\mathcal{A}}
\newcommand{\cL}{\mathcal{L}}
\newcommand{\cN}{\mathcal{N}}
\newcommand{\cG}{\mathcal{G}}
\newcommand{\cH}{\mathcal{H}}
\newcommand{\cV}{\mathcal{V}}
\def\E{{$E_{7(7)}$}}

\def\K{{K\"{a}hler}}

\renewcommand{\a}{\alpha}

\def\vp{\varphi}

\newcommand{\bea}{\begin{eqnarray}}
\newcommand{\eea}{\end{eqnarray}}
\newcommand{\be}{\begin{equation}}
\newcommand{\ee}{\end{equation}}
\newcommand{\rf}[1]{(\ref{#1})}

\usepackage{cite}

\begin{document}

\setcounter{footnote}{0}

\baselineskip 6 mm

\begin{titlepage}
	\thispagestyle{empty}

{\ } 
	
	\vspace{35pt}
	
	\begin{center}
	    { \huge{\bf Attractors in Supergravity  }}
		
		\vspace{50pt}
		
		{\bf Renata Kallosh and Andrei Linde}
		
		\vspace{25pt}

		  {\it Stanford Institute for Theoretical Physics and Department of Physics,\\ Stanford University, Stanford, CA 94305, USA}
		
		\vspace{15pt}

		\vspace{40pt}
		
		{ABSTRACT}
	\end{center}

The concept of attractors, well-known in classical mechanics, proved to be very productive in supergravity, in the 
theory of black holes and inflationary cosmology. We start with attractors in supersymmetric black holes and discuss also non-BPS black hole attractors. Recently the non-BPS case helped to explain, via enhanced dualitiy symmetry, mysterious cancellation of ultraviolet divergences in 82 Feynman diagrams in 4-loop superamplitude in $\cN=5$ supergravity.  We discuss the implications of these results for the possibility of the all-loop finiteness of $N >4$ 4D supergravities.

 We continue with the description of inflationary $\alpha$-attractors. This large class of inflationary models gives predictions that are stable with respect to even very significant modifications of inflationary potentials. These predictions match all presently available CMB-related cosmological data. These models provide targets for the future satellite mission LiteBIRD, which will attempt to detect primordial gravitational waves. We describe some of the recent advanced versions of cosmological attractors which have a beautiful fractal landscape structure.

\vspace{30pt}

\begin{center}
\emph{Invited contribution to ``Half a century of Supergravity'' \\ eds.~A. Ceresole and G.~Dall'Agata (Cambridge Univ. Press, to appear)}
\end{center}

\bigskip

\end{titlepage}

\newpage

\tableofcontents

\newpage

\parskip 3pt 

\section{\sc Introduction}

\begin{figure}[h!]
\centering
\includegraphics[width=150mm]{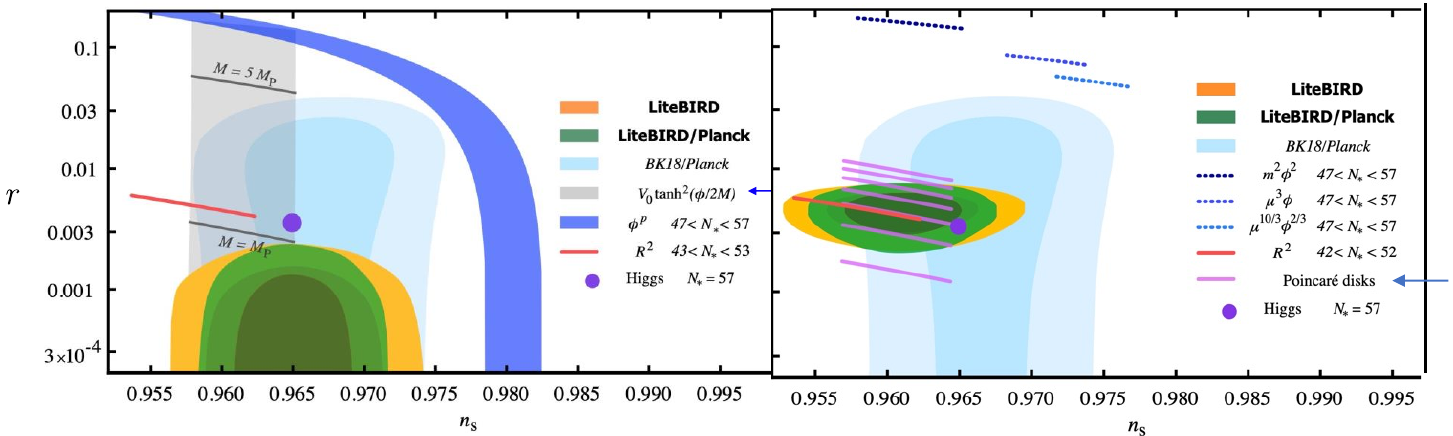}
\caption{\small LiteBIRD  satellite mission CMB targets  in $n_s-r$ plane \cite{LiteBIRD:2022cnt}.  Supergravity targets include simplest $\alpha$-attractor cosmological models of inflation. On the left panel, there is a grey band \cite{Kallosh:2013yoa} with a potential $\tanh^2(\varphi/\sqrt{6\alpha}) $. On the right panel, there are 7 Poincar\' e disks from string theory inspired supergravity $\alpha$-attractor  inflationary models \cite{Ferrara:2016fwe}. The launch date of LiteBIRD is expected in Japanese fiscal year 2032. 
}
\label{fig:LiteBIRD}
\end{figure}

 A cosmological setting presents a rare situation where the theoretical predictions of supergravity are tested experimentally. In particular, supergravity-based inflationary $\alpha$-attractor models  \cite{Kallosh:2013yoa} have already been tested by Planck \cite{Planck:2018jri}  and by BICEP/Keck \cite{BICEP:2021xfz}, and will be tested even further by future cosmological observations  \cite{LiteBIRD:2022cnt}. We show in Fig. \ref{fig:LiteBIRD} supergravity targets which are predictions of the simplest $\alpha$-attractor cosmological models of inflation  \cite{Kallosh:2013yoa,Ferrara:2016fwe}.
Cosmic Microwave Background community is well aware of supergravity targets as one can see from Fig. \ref{fig:LiteBIRD} : these targets are taken from  ``Probing cosmic inflation with the LiteBIRD cosmic microwave background polarization survey'' produced by the LiteBIRD collaboration  \cite{LiteBIRD:2022cnt}.

The study of attractors in supergravity began about two decades earlier, with an investigation of extremal black hole solutions in supergravity, long before the cosmological attractor models of inflation were constructed. We refer here to a review of black hole solutions in theories of supergravity
in this book by T. Ortin  \cite{Ortin:2024slu}.

\begin{figure}[h!]
\centering
\includegraphics[width=60mm]{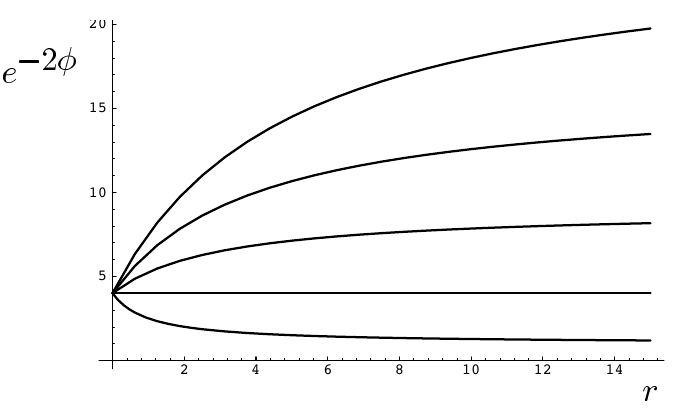}
\caption{\small A very first explicit example of the supersymmetric black hole attractor \cite{Ferrara:1995ih}.  All values of the dilaton $e^{-2\phi}(r)$ away from the horizon at large $r$ are different; however, near the horizon, they all go to an attractor point defined by the black hole charges. }
\label{fig:BHattractor}
\end{figure}

The black hole attractor story in supergravity started at a cafe in Aspen in Summer 1995 when Sergio Ferrara, Andy Strominger and one of the authors met and started talking about supergravity black holes and teaching classical mechanics. It was soon realized in  \cite{Ferrara:1995ih}  that in the case of black holes, we have discovered an attractor with the evolution parameter, which is not time (as in most examples in non-linear dynamics) but a distance to the horizon, see Fig. \ref{fig:BHattractor}.

A simple example of an attractor behavior of the dilaton in Fig. \ref{fig:BHattractor} is given by a solution of Einstein equations for the metric and two vectors with electric and magnetic charges, and a dilaton, where the dilaton depends on the distance to the horizon $r$ as follows
\be
e^{-2\phi} = {e^{-\phi_0} +{|q|\over r}\over e^{\phi_0} +{|p|\over r}} \ .
\ee
Far away from the horizon, at $r\to \infty$ $e^{-2\phi}\to e^{-2\phi_0}$, near the horizon, at $r\to 0$,  $e^{-2\phi}\to {|q|\over |p|}$. The value of the dilaton $e^{-2\phi}$ near the horizon ${|q|\over |p|}$ is universal and independent on initial conditions $e^{-2\phi_0}$.

The basic upshot of more general cases of supergravity black hole  attractors in ungauged supergravity is  that the unbroken supersymmetry is enhanced near the horizon, all memory about initial conditions far away from the horizon is lost, and there is a universal near horizon 
 $AdS_2\times S^2$  geometry. 
 
 Here we will focus on $\cN=8$ non-BPS black hole attractors \cite{Ferrara:2006em,Ceresole:2009jc}, which have recently attracted attention to the fact that in each dimension D, there are type I and type II ungauged supergravities \cite{Kallosh:2024rdr}. Type I is the well-known standard supergravity with ${\cG\over \cH}$ coset space for each dimension D, see also H. Nicolai review in this book \cite{Nicolai:2024hqh}. A detailed  procedure of dualization of supergravities starting from 11D, which leads to ${\cG\over \cH}$ coset space for each D, is described in 
 \cite{Cremmer:1997ct}.
 Type II supergravities in dimension D are obtained by compactifying supergravities in higher dimensions, $D+n$, but without dualization. It is important to stress here that all symmetries $\cG$ and $\cH$ in type I supergravities are achieved only after dualization. At the classical level, dualization might relate to each other equivalent theories. However, this equivalence at the quantum level is the issue here. It is relevant for understanding superamplitude loop computations in supergravity.

 Consider, for example,  maximal supergravity in 4D  \cite{Cremmer:1979up}. The 70 scalars with non-polynomial interaction in  $\cN=8$ supergravity are in a coset
$
{\cG\over \cH}= {E_{7(7)}\over SU(8)}
$.
An example of the maximal supergravity in 4D  of type II  is the one in \cite{Andrianopoli:2002mf} which is a 5D supergravity compactified to 4D on a circle \footnote{The supergravity in  \cite{Andrianopoli:2002mf} can also be obtained from the gauged supergravity \cite{Sezgin:1981ac} in the limit of vanishing gaugings.}. The 70 scalars are split into 42 in the 5D coset ${\cG\over \cH}= {E_{6(6)}\over USp(8)}$, and a radius of a circle, all have a non-polynomial interaction, and there are 27 axions with polynomial interaction. Also, in type I supergravity  \cite{Cremmer:1979up}, there are  28 doublet vectors of \E\,  and they are split into 27 doublet vectors of $E_{6(6)}$ and a single doublet vector in \cite{Andrianopoli:2002mf}. Classically, type II supergravity, upon dualization, will acquire the $SU(8)$ and \E\  symmetries, but at the quantum level, these duality transformations may or may not give equivalent S-matrices.

It was discovered in \cite{Ceresole:2009jc} that the extremal non-BPS  Kaluza-Klein black holes have a natural embedding into type II supergravity, whereas the 1/8 BPS are embedded into type I supergravity. 

This distinction between two types of supergravity in the same dimension with the same amount of supersymmetries, which was necessary to understand the difference between BPS and non-BPS black hole attractors, led us to question the quantum equivalence of these supergravities. This, in turn, required to define the concept of ``enhanced dualities'' \cite{Kallosh:2024ull}, which helped to explain ``enhanced 
cancellations'' of UV infinities in 82 diagrams in $\cN=5$ at loop order 4 \cite{Bern:2014sna,Bern:2023zkg}.

\

\begin{figure}[t]
\centering
\includegraphics[width=95mm]{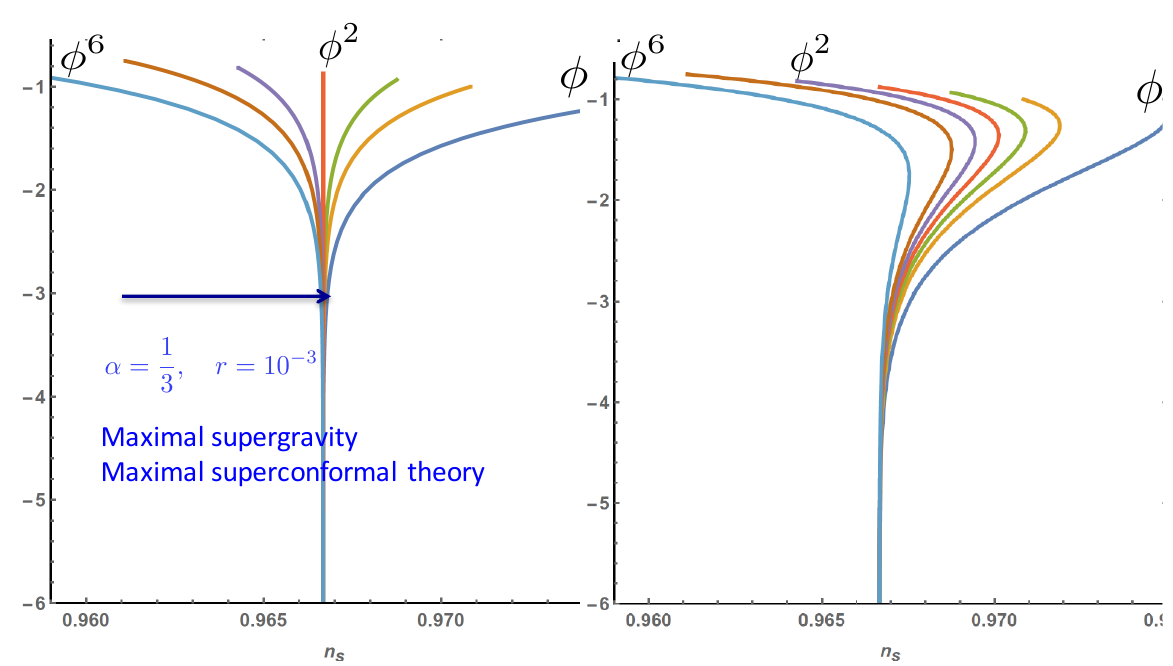}
\caption{\small The predictions of cosmological T-models on the left and E-models on the right for  $r$ versus $n_s$ CMB observables, $r$ is in a log scale. In these models $n_s$ is $\alpha$-independent, whereas $r={12\alpha \over N^2}$. For a fixed number of e-foldings $N$ and decreasing $\alpha$, the predictions of models $\phi^n$ converge to $n$-independent attractor values.
}
\label{fig:CosmAttractors}
\end{figure}

The story of cosmological attractors started for us in Summer 2013 when we were driving from Stanford to Santa Barbara and made a stop halfway at a burger place, which was famous at that time. The waiting line was very long. We started writing some equations using their paper napkins, and we finished the calculations in the evening when we arrived to Santa Barbara. We have found, see    \cite {Kallosh:2013hoa},  that at $\a=1$ inflationary predictions for the potential depending on $\tanh^{2n}\vp$ are $n$-independent.   We have soon constructed models in \cite{Kallosh:2013yoa} with arbitrary $\a$, and we found that that the evolution parameter is given by $ {3\alpha\over 8N}$ where $R_K=-{2\over 3\alpha}$ is the Kahler curvature of the field theory space, formed by the inflaton and axion, and $N$ is the number of e-foldings of inflation. It is usually taken to be of the order of 55, whereas $\a$ can change significantly. We show the attractor properties at decreasing $\a$ of various inflationary models in   Fig. \ref{fig:CosmAttractors}.

The predictions of the $\alpha$-attractor models \cite{Kallosh:2013yoa} for observable $n_s$ and $r$ were in agreement with just released at that time observations of Planck 2013 and still in agreement with Planck 2018 and BICEP/Keck 2021  \cite{Planck:2018jri,BICEP:2021xfz}.  They remain targets for the future satellite CMB mission LiteBIRD \cite{LiteBIRD:2022cnt} as we show in Fig. \ref{fig:LiteBIRD}.  Recent data from the South Pole Telescope \cite{SPT-3G:2024atg}, combined with Planck and WMAP data, see Table IV there,  shows that the tilt of the spectrum  $ n_s=0.9647 \pm 0.0037$,  from Planck and SPT, or smaller with WMAP,  is in agreement with  $\alpha$-attractor models. 
 Same in \cite{Galloni:2022mok} where the data from BICEP/Keck Array  2018, Planck21, and LIGO-Virgo-KAGRA Collaboration is given by $ n_s=0.9676 \pm  0.0039$.

We have proposed recently a general case of $SL(2, \mathbb{Z})$ cosmological $\alpha$-attractors \cite{Kallosh:2024ymt,Kallosh:2024pat,Kallosh:2024whb}, following the idea in \cite{Casas:2024jbw} that one can take inspiration from string theory, in using $SL(2, \mathbb{Z})$  symmetry and combine it with $SL(2, \mathbb{Z})$-invariant {\it plateau} potential. Note that the $SL(2, \mathbb{Z})$-invariant plateau potentials compatible with cosmological observation have not been derived from string theory. But these models realize space-time target space duality \cite{Ferrara:1989bc}. The idea in \cite{Ferrara:1989bc} was that in the $SL(2,\mathbb{R})$ symmetric supergravity, one can expect that continuous symmetry is broken by non-perturbative effects including instantons. However, it is possible that a potential with modular invariance may not be subject to quantum corrections if modular symmetry is an exact symmetry. It is, therefore, interesting to find out if general type $SL(2, \mathbb{Z})$-invariant {\it plateau} potentials, which we describe, have universal predictions for cosmological observables.

Target space duality is different from duality symmetry in string theory since we study physics in 4D space-time, not on the 2D world-sheet. Here we refer the reader to a  contribution to this book by Cribiori and  Lust on 
 string dualities and modular symmetries in supergravity \cite{Cribiori:2024qsv}.

The reason why, in the past, it was difficult to construct supergravity modular inflation models is that in traditional cases, one had to 
find a modular invariant action starting from a  \K\, invariant supergravity function of the form
$
\mathcal G(T, \bar T)= K(T, \bar T) + \log |W(T)|^2
$.

Meanwhile, it became clear, starting with KKLT anti-D3 brane uplift in string theory \cite{Kachru:2003aw}, that cosmology in the presence of a nilpotent superfield in supergravity and a Volkov-Akulov type nonlinearly realized supersymmetry could produce supergravity models compatible with the observations.
See the review of non-linear supergravity and inflationary cosmology in this book by Antoniadis,  Dudas,  Farakos, and Sagnotti in \cite{Antoniadis:2024hvw}.  Specifically, we will use the geometric construction of supergravity presented in 
\cite{Kallosh:2017wnt}, \cite{Kallosh:2022vha} which was  applied to $SL(2,\mathbb{Z})$ cosmology in \cite{Kallosh:2024ymt}.

The advantage of the nilpotent superfield $X$ for constructing de Sitter supergravity  
 and modular invariant cosmological models in supergravity will be demonstrated in Sec. \ref{Sec:dS} for dS supergravity \cite{Bergshoeff:2015tra} and in 
\ref{nonlinsuper} in the context of $SL(2,\mathbb{Z})$ models \cite{Kallosh:2024ymt}.  We will show there that the requirement of supersymmetry does not constrain the parameter $\a$ in $SL(2,\mathbb{Z})$ supergravity models once we use the unitary gauge for local supersymmetry. In the unitary gauge, the second derivative of the \K\,  potential over the nilpotent superfield $G_{X\bar X}(T, \bar T)$ is not required to be positive definite.

\section{ Black hole attractors}
\subsection{$\cN=8$ attractors, BPS and non-BPS}
Since the studies of black hole attractors in supergravity started a few decades ago, there is a fair amount of information in various lectures, for example,  \cite{Moore:1998zu}, in the supergravity textbook \cite{Freedman:2012zz} and in the contribution to this book in  \cite{Ortin:2024slu}. In \cite{Freedman:2012zz}, the attractor mechanism is presented first as ``slow and simple'' and proceeds in a ``fast and furious'' way,  up to the case of general $\cN=2$ supergravity coupled to $n$ abelian vector multiplets. The concepts of 1/2 BPS extremal black holes with unbroken 1/2 of supersymmetries as well as black hole potential and its critical points,  are well explained there.
We, therefore, will just add here that non-BPS extremal black holes with all supersymmetries broken in $\cN=2$ supergravity are also well known, starting with  \cite{Goldstein:2005hq}.   It is shown in  \cite{Ceresole:2007wx} that stable extremal non-BPS black holes can be described by first-order differential equations driven by a ``superpotential'',  replacing central charge in the usual black hole potential.

A recent review of $\cN = 2$ supergravity and its classic attractor mechanism, as well as  
the counting of microstates for supersymmetric black holes obtained from a supersymmetric index in weakly-coupled string theory, is presented in \cite{Boruch:2023gfn}. This brings a context in which more recent studies of BPS black holes were performed. Currently, this direction remains interesting and developing, see also Cassani and Murthy contribution to this book in \cite{Cassani:2025sim}.

We will proceed here with the discussion of regular horizon $\cN=8$ BPS and non-BPS black hole attractors \cite{Ferrara:2006em,Ceresole:2009jc} based on classical $\cN=8$ ungauged supergravity \cite{Cremmer:1979up}. The scalar manifold in $\cN=8$ supergravity is the coset
$
{\cG\over \cH}= {E_{7(7)}\over SU(8)}
$.
As we will see, this is one of many aspects of black hole attractors in supergravity, which is close to the frontiers of the current theoretical physics based on superamplitudes.

Recent advances in superamplitudes computations described in \cite{Bern:2023zkg} show that 4D $\cN\geq 5$ supergravities have an unexpected UV behavior. The most interesting case is one of the so-called ``enhanced cancellation'' of UV divergences at loop order $L=4$ in $\cN=5$ supergravity, where UV infinities in 82 diagrams cancel \cite{Bern:2014sna}.

 $\cN=8$ attractors  \cite{Ferrara:2006em} were identified using the standard strategy of finding critical points of the corresponding black hole potential, in full analogy with the $\cN=2$ case. In the $\cN=8$ case, the derivation of all critical points is actually simple! 
 
The black hole entropy for $\cN=8$ supergravity was known long before the attractor mechanism for $\cN=8$ supergravity was described in  \cite{Ferrara:2006em,Ceresole:2009jc}. Namely, it was shown in   \cite{Kallosh:1996uy}  from  U-duality,  that  the 1/8 BPS black hole entropy is given by a  Cartan-Cremmer-Julia quartic \E\,  invariant $J_4$ depending on black hole electric and magnetic charges in the fundamental {\bf 56} representation of \E\, 
 \be
{S_{BPS}\over \pi} = \sqrt {J_4(p,q)}\ ,  \qquad J_4>0 \ .
\label{BPS1}\ee 
A  symplectic charge matrix-vector  $Q$ for $\cN=8$ consists of electric $q\to e_{\Lambda \Sigma}$ and magnetic $p\to m^{\Lambda \Sigma}$ charges forming the fundamental representation of $E_{7(7)}$
\be
Q \equiv (m^{\Lambda \Sigma}, e_{\Lambda \Sigma} ) \ .
\ee
 A scalar-dependent symplectic doublet and its conjugate are introduced as follows 
\be
V_{AB} = \begin{pmatrix}
 f_{AB}^{\Lambda \Sigma}\\
h_{\Lambda \Sigma, AB }
\end{pmatrix} \ ,
  \qquad \bar V^{AB} = \begin{pmatrix}
 \bar f^{\Lambda \Sigma, AB}\\
h_{\Lambda \Sigma}^{ AB }
\end{pmatrix}  \ .
\ee
Here the pair of indices $\Lambda\Sigma$ in  $f_{AB}^{\Lambda \Sigma}= - f_{AB}^{\Sigma \Lambda}$   run over the $\mathbf{28}$ of $SL(8, \mathbb{R})$ and in $\mathbf{28}'$ in $h_{\Lambda \Sigma, AB }$. The pair of indices $AB$ in $f_{AB}^{\Lambda \Sigma}= - f_{BA}^{\Lambda \Sigma}$, run over the $\mathbf{28}$ of $SU(8)$  for $f$ and $h$ but in $\overline{ \mathbf{28}}$ for the $\bar f$ and $\bar h$.
A symplectic invariant  central charge matrix and its conjugate are 
\be
Z_{AB}= f_{AB}^{\Lambda \Sigma}e_{\Lambda \Sigma} - h_{\Lambda \Sigma, AB } m^{\Lambda \Sigma}
\equiv \langle Q, V_{AB} \rangle \, , \qquad Z^{*AB}= \langle Q, \bar V^{AB} \rangle
\ee
The  black hole potential in $\cN=8$,  4D supergravity  is
\be
{\cal V} _{BH}(\phi, Q) = Z_{AB}  Z^{*AB} = \langle Q,  V_{AB} \rangle \langle Q, \bar V^{AB} \rangle \qquad A,B=1,\dots , 8.
\label{N8pot}\ee
The covariant derivative of the central charge  is defined by the Maurer-Cartan equations for the coset space:
$
{\mathcal D}_i Z_{AB}={1\over 2} P_{i,[ ABCD]}(\phi)  Z^{* CD}(\phi, Q)
$.
Here 
$
P_{i,[ ABCD]}= {1\over 4!} \epsilon_{ABCDEFGH}(P_{i}^{*[ EFGH] })
$ and 
$D_i$ is the SU(8) covariant derivative.
Thus the derivative of the black hole potential over 70 moduli is given by the following expression
\be
\partial_i {\cal V}={1\over 4}  P_{i,[ ABCD]}\Big [Z^{*[CD} Z^{*AB]} + {1\over 4!} \epsilon ^{CDABEFGH} Z_{EF} Z_{GH} \Big ]
\label{critical1}\ee
The $70\times 70$-bein $P_{i,[ ABCD]}$ is invertible. Therefore a necessary and sufficient condition defining  the critical points of the black hole potential with regular $70\times 70$-beins is an {\it algebraic} \footnote{It is interesting to compare it with $\cN=2$ case where  $Z(z, \bar z)= 
(L^\Lambda q_\Lambda  - M_{\Lambda  } p^{\Lambda})
\equiv \langle Q, V \rangle$ and  
$D_i Z= (\partial_i +{1/2} K_i) Z(z, \bar z, p, q)$ implies that ${\partial \over \partial z^i} |Z|=0$. This {\it differential} equation is solved in the form $p^\lambda = i(\bar Z L^\Lambda- Z \bar L^\Lambda)$, $q_\Lambda = i(\bar Z M_\Lambda- Z \bar M_\Lambda)$
so that the attractor values of scalars $z, \bar z$ become functions of charges $p, q$.}
condition:
\be
Z^{*[AB}Z^{*CD]}+ {1\over 4!} \epsilon ^{ABCDEFGH} Z_{EF} Z_{GH}=0
\label{N8}\ee
It is a condition extremizing the black hole potential.
 The antisymmetric central charge matrix  has  four  non-vanishing complex eigenvalues $z_1=Z_{12}, \, z_2=Z_{34}, \, z_3= Z_{56}, \, z_4=Z_{78}$. In this basis, the attractor equations  are
\bea
 z_1 z_2 + z^{*3}z^{*4}=0\ , \qquad 
 z_1 z_3 + z^{*2}z^{*4}=0\ , \qquad 
 z_2 z_3 + z^{*1}z^{*4}=0 \ .
\label{attractors}\eea
The $SU(8)$ symmetry allows bringing all 4 complex eigenvalues to the following normal form \cite{Ferrara:1997ci}
\be
z_i= \rho_i e^{i\varphi/4} \ , \qquad i=1,2,3,4.
\label{basis}\ee
Only 5 real parameters are independent, 4 absolute values $\rho_i$ and an overall phase, $\varphi$,  since the relative phase of each eigenvalue can be changed but not the overall phase.

The quartic $J_4$ invariant can be given as a function of central charges
\be
J_4= Tr (Z\bar Z)^2 -{1\over 4} (Tr Z\bar Z)^2+ 4 (PfZ +Pf\bar Z) \ .
\ee
In the basis \rf{basis} it acquires the following form  \cite{Ferrara:1997ci}
\be
J_4=\Big [ (\rho_1+ \rho_2)^2- (\rho_3+ \rho_4)^2\Big] \Big [ (\rho_1- \rho_2)^2- (\rho_3- \rho_4)^2\Big]+ 8 \rho_1 \rho_2 \rho_3 \rho_4 (\cos \varphi -1)
\ee
$\cN=8$ attractor equations (\ref{attractors}) have 2 solutions for regular black holes 
\begin{enumerate}
  \item 1/8 BPS solution 
  \be
  z_1= \rho_{BPS} e^{i\varphi_1} \neq 0\, ,  \qquad z_2=z_3=z_4=0\, ,   \qquad J_4^{BPS}=\rho_{BPS}^4>0 \ .
  \ee
  The black hole entropy and the area of the horizon   of the BPS black holes with 1/8 of $\cN=8$ unbroken supersymmetry is given by
\be
{S_{BPS}(Q)\over \pi} = {A_{BPS}(Q)\over 4 \pi}= \sqrt {J_4^{BPS}(Q)}= \rho_{BPS}^2
\label{BPS}\ee
The quartic invariant is positive. The 1/8 BPS solution  breaks the $SU(8)$ symmetry 
\be
SU(8) \to SU(2) \times U(6)
\ee
  \item non-BPS solution
  \be
z_i= \rho_{nonBPS} \, e^{i{\pi\over 4}}\, ,   \qquad J_4^{nonBPS}= -16 \rho_{nonBPS}^4 <0
  \ee
    The black hole entropy and area formula of the non-BPS black holes, with all supersymmetries broken, is given by
\be
{S_{nonBPS}(Q)\over \pi} = {A_{nonBPS}(Q)\over 4 \pi}= \sqrt {-J_4^{BPS}(Q)}= 4 \rho_{nonBPS}^2
\label{nonBPS}\ee
The quartic invariant is negative. The non-BPS solution  breaks the $SU(8)$ symmetry  
\be
SU(8) \to USp(8) 
\ee
\end{enumerate}
Soon after the discovery of the non-BPS critical points of the potential  in $\cN=8$ supergravity, it has been realized in 
 \cite{Ceresole:2009jc} that the non-BPS  Kaluza-Klein black holes have a natural embedding in type II supergravity \cite{Andrianopoli:2002mf}, whereas the 1/8 BPS are embedded into type I supergravity \cite{Cremmer:1979up}. 

\subsection {Ungauged Supergravities of type I and type II}
We refer to a general description of type I and type II supergravities to \cite{Kallosh:2024rdr}. Here we will focus on 4D examples, which are most important both in the context of non-BPS black hole attractors as well as in the case of enhanced amplitude cancellations in \cite{Bern:2014sna} of  82 diagrams in $\cN=5$ at loop order 4,  
and enhanced dualities 
\cite{Kallosh:2024ull} which  explain this enhanced 
cancellation of UV infinities. 

Supergravity actions depend on  scalars via the vielbein $\cV(x)$.
The vielbein transforms under global $\cG$ symmetry and local ${\cH}$-symmetry
\be
\cV(x) \to {\bf g}\, \cV(x)h^{-1}(x)
\ee
Before gauge-fixing local $\cH$ symmetry, the vielbein is in the adjoint representation of ${\cG}$, and the number of scalars is dim [${\cG}$]. After gauge-fixing
$
\cV(x)\to \cV(x)_{g.f.}
$ and 
it is a matrix depending  only on  physical scalars, where the number of physical scalars  is equal to
dim [${\cG}$] - dim [$\cH$].

For example, the $D=4$, $N=8$ supergravity with manifest $\cG$=\E\,  global
symmetry was obtained by dimensionally reducing
11D supergravity, and then dualizing the seven 2-form
potentials to give seven scalars, and dualizing twenty-one
pseudo-vectors to give twenty-one vectors. These twenty-one vectors, together with
the seven Kaluza-Klein vectors, form a 28-dimensional
representation of $SL(8, \mathbb{R})$.  The final 4D supergravity type I action in
\cite{Cremmer:1979up} has  70 scalars with non-polynomial interaction in  $\cN=8$ supergravity  in a coset
$
{\cG\over \cH}= {E_{7(7)}\over SU(8)}
$.

In type I  supergravity action \cite{Cremmer:1979up} the pure scalar part, before gauge-fixing local SU(8) symmetry has the form
\be
{1\over e} \cL^{I\, sc}_{_{ 4D}} = {1\over 4!} P_\mu{}^{ijkl} P^\mu{}_{ijkl} \ ,
\ee
where $
({\cV}^{-1} D_\mu \cV)^{ijkl} = {\cal P}_\mu^{ijkl}
$ is a local SU(8) tensor in ${\bf 70}$. The scalar-vector Lagrangian in a symmetric gauge in 4D has the form
 \begin{equation}
\frac{1}{e}{\cal L}^{I\, vec}_{_{ 4D}} =
\frac{1}{4}\, {\cal I}_{IJ}(\phi)\,F^I_{\mu\nu}\,F^{J\,\mu\nu}
+\frac{1}{8\,e}\,{\cal R}_{IJ}(\phi)\,\epsilon^{\mu\nu\rho\sigma}\,F^I_{\mu\nu} \,F^{J}_{\rho\sigma}\ .
\label{boslagr2}
\end{equation}
 Here $I,J =1,\dots , 28$. The vector couplings ${\cal I}_{IJ}(\phi), \, {\cal R}_{IJ}(\phi)$ depend non-polynomially on 70  self-dual scalars $\phi_{ijkl}= \pm {1\over 4!} \epsilon_{ijklpqmn}\bar \phi^{pqmn}$ which transform in the 35-dimensional representation of $SU(8)$.

The scalar action in 4D  type II  supergravity  in \cite{Andrianopoli:2002mf}, used in the context of  non-BPS black hole attractors in \cite{Ceresole:2009jc}
 is
\be
{1\over e}\,\cL^{II\, sc}_{_{ 4D}} = {3\over 2} \partial_\mu \phi \partial^\mu \phi -{1\over 4} e^{-4\phi} \hat {\cal N} _{\Lambda \Sigma}  \partial_\mu a^\Lambda  \partial^\mu a^\Sigma +{1\over 4!} P_\mu{}^{abcd} P^\mu{}_{abcd} \ .
\ee
Here $\hat {\cal N} _{\Lambda \Sigma}$ is the 5D ($SO(1,1)$ invariant) vector kinetic matrix, $\Lambda=1,\dots 27$,  $P_\mu{}^{abcd}=
({\cV}^{-1} D_\mu \cV)^{abcd}$ depends on a 5D ${E_{6(6)}/USp(8)}$ vielbein and is a local $USp(8)$ tensor in ${\bf 42}$.
This corresponds to a decomposition of 70 scalars under $USp(8)$ as ${\bf 70}\to {\bf 1}+{\bf 27}+{\bf 42}$. It means there are 42 scalars from the 5D coset ${E_{6(6)}/USp(8)}$, one scalar, the radius of the circle, and 27 axions.

 The 28 vectors in the action in \cite{Cremmer:1979up} in \cite{Andrianopoli:2002mf} are represented by 1+27 vectors in  \cite{Andrianopoli:2002mf,Ceresole:2009jc}
$
B_\mu\, , \, Z_\mu^\Lambda
$.
 Both actions depend only on field strength's: 28 in  $F_{\mu\nu}^{IJ} =\partial_\mu \cA_\nu^{IJ} -\partial_\nu \cA_\mu^{IJ} \,$ in \cite{Cremmer:1979up} and 1+27 $B_{\mu\nu} =\partial_\mu B_\nu-\partial_\nu B_\mu\,$ and $Z_{\mu\nu}^\Lambda =\partial_\mu B_\nu^\Lambda-\partial_\nu B_\mu^\Lambda\,$ in  \cite{Andrianopoli:2002mf}.
 The scalar-vector Lagrangian is
\bea
&&\frac{1}{e}{\cal L}^{II \, vec}_{4D}=
 {\cal I}_{00}(\phi)\,B_{\mu\nu} B^{\rho\sigma}+ 2 {\cal I}_{0\Lambda}(\phi)\,B_{\mu\nu}\,Z^{\Lambda\,\mu\nu}+{\cal I}_{\Lambda\Sigma}(\phi)\,Z^\Lambda_{\mu\nu}\,Z^{\Sigma\,\mu\nu}
\cr
\cr
&+&\frac{1}{2\,e} \epsilon^{\mu\nu\rho\sigma}  [ {\cal R}_{00}(\phi) B_{\mu\nu} B_{\rho\sigma}  +2{\cal R}_{0\Lambda}(\phi) B_{\mu\nu} \,Z^{\Lambda}_{\rho\sigma}+{\cal R}_{\Lambda \Sigma}(\phi) Z^\Lambda_{\mu\nu} \,Z^{\Sigma}_{\rho\sigma}]
\label{boslagr}
\eea
where ${\cal I}_{IJ}$ and ${\cal R}_{IJ}$ are given by 4 blocks with 28 split into 0 and 27 $\Lambda$'s.
These depend on  $d_{\Lambda \Sigma \Gamma}$, 
a symmetric invariant tensor of the representation {\bf 27} of $E_{6(6)}$ and $a_{\Lambda \Sigma}$, a five-dimensional
SO(1,1) invariant vector kinetic matrix. The
 scalar-dependent kinetic terms of vectors
 are polynomial in axions $a^{\Lambda}$.
 Both actions in 4D, supergravity I and II, have maximal local supersymmetry when supplemented with fermions. Supergravity II has inherited local supersymmetry via dimensional reduction. 

By comparing the scalar and vector actions in maximal 4D supergravity of type I and of type II, it is not obvious, even at the classical level, that these describe equivalent theories. Fortunately, in 4D, there is a Gaillard-Zumino (GZ) electro-magnetic symmetry  \cite{Gaillard:1981rj}. Its role in relating the 4D supergravity of type I to type II  was revealed by de Wit, Samtleben, and Trigiante (dWST) in \cite{deWit:2002vt}.

\subsection{Enhanced dualities explaining superamplitude  computations}
\begin{figure}[t]
\centering
\includegraphics[width=105mm]{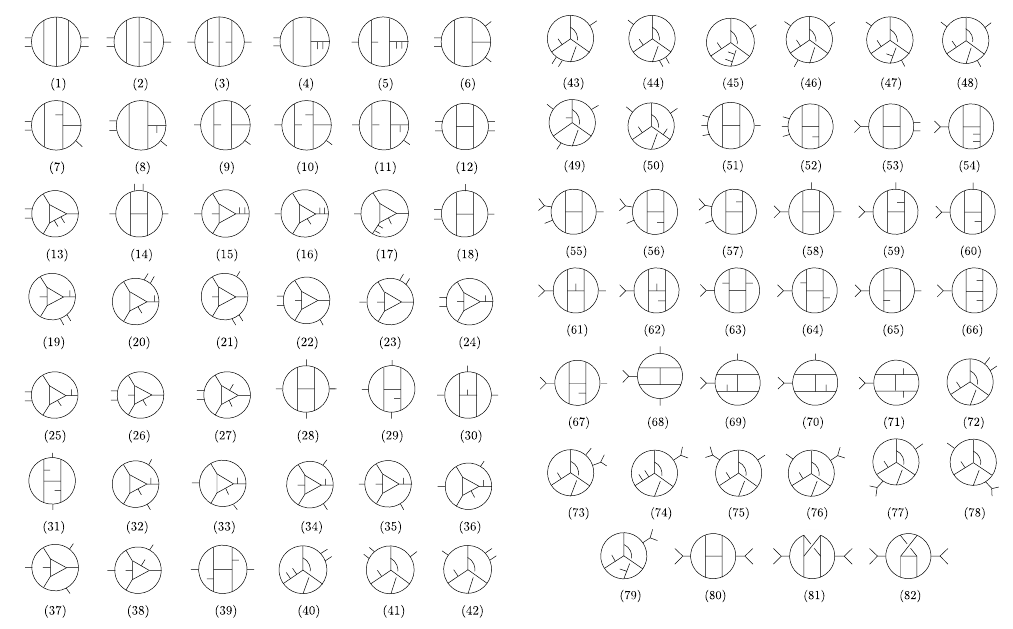}
\caption{\small 82 diagrams in 4-loop superamplitude in $\cN=5$ supergravity computed in \cite{Bern:2014sna}. Each diagram is UV divergent but the sum is UV finite.}
\label{fig:82_N5}
\end{figure}

U-duality imposes constraints on the structure of divergences in supergravity. But GZ  Sp$(2n_v, \mathbb{R})$ duality in 4D has more symmetries than U-duality.  For example, in $\cN = 8$, the dimension of Sp(56) is 1596, whereas its U-duality subgroup  \E\ has dimension 133. In comparison, in  D $> 4$, maximal dualities are U-dualities. 
In this section, we will present a short summary of the results on the role of GZ symmetry  \cite{Gaillard:1981rj} in the dWST construction 
\cite{deWit:2002vt} and  the argument in \cite{Kallosh:2024ull} explaining the superamplitude computations in 
\cite{Bern:2014sna}. 

There was also an earlier prediction in \cite{Bossard:2011tq}, based on harmonic superspace counterterms,  that the 4-particle scattering amplitude at loop order $L=\cN-1$ will be UV divergent. This prediction was invalidated for the case $L=4$, $\cN=5$ by computations in \cite{Bern:2014sna}, where the UV  divergences in 82 diagrams canceled, see Fig. \ref{fig:82_N5}.  More recently, in \cite{Bern:2023zkg}, this cancellation of UV divergences was qualified as  an example of a ``puzzling enhanced ultraviolet cancellations, for which no symmetry-based understanding currently exists.''

We argue   in \cite{Kallosh:2024ull} that the extra dualities in 4D, enhancing U-duality, determine the properties of perturbative quantum supergravity.  The presence/absence of enhanced dualities suggests a possible explanation of the results of the amplitude loop computations in D-dimensional supergravities and of the special status of 4D  in this respect.

In 4D $\cN\geq 5$, GZ duality group is 
\be\boxed{
Sp(2n_v, \mathbb{R})\supset \cG_U } \ ,
\ee
 whereas U-duality group 
$
\cG_U$ is a subgroup of $Sp(2n_v, \mathbb{R}) $.
 In particular for $\cN\geq 5$ there is a global  GZ duality symmetry, $\cG_{GZ}$, a global U-duality $\cG$, and a local symmetry $\cH$
\bea
&&\cN=8:  \, \,   \cG_{GZ}=Sp(56,  \mathbb{R}) \supset E_{7(7)} \qquad \quad \, \,  {\cG\over \cH}= {E_{7(7)}\over SU(8)}\cr
\cr
&&  \cN=6:  \, \,  \cG_{GZ}=Sp(32,  \mathbb{R})  \supset SO^*(12)  \qquad {\cG\over \cH}= {SO^*(12)\over U(6)}\cr
\cr
&&  \cN=5: \, \,  \cG_{GZ}= Sp(20,  \mathbb{R})  \supset SU(1,5) \qquad {\cG\over \cH}= {SU(1,5)\over U(5)}
\label{N568}\eea
The main feature of the dWST construction 
\cite{deWit:2002vt} is that there are different symplectic frames in ungauged 4D $\cN\geq 5$ supergravities. There are symmetries presented as a  double  quotient
\be
E^{{4D}}=\cG_U (\mathbb{R})  \backslash Sp(2n_v; \mathbb{R})/GL(2n_v)
\label{DQ}\ee
which relate supergravities of type I and type II and allowed to prove that these are classically equivalent. These double quotients are non-trivial in all 4D $\cN\geq 5$ supergravities since the GZ duality group is bigger than the U-duality group in each of these cases.

The dWST construction 
\cite{deWit:2002vt} was uplifted to a quantum level in \cite{Kallosh:2024ull} using GZ duality symmetry in the path integral and the 
Hamiltonian formulation of dualities developed in \cite{Henneaux:2017kbx}. Therefore a bona fide $Sp(2n_v,R)$ Noether current can be constructed, which indicates that GZ duality symmetry (or rather the elements of it outside $\cG_U $ and 
$GL(2n_v)$  which we called ``enhanced dualities") protect 4D $\cN\geq 5$ from UV divergences. This is a symmetry-based explanation of the cancellation of 82 diagrams displayed in Fig. \ref{fig:82_N5}, see also https://www.ias.edu/sns/amplitudes-2024-scientific-program.
We view this result as an indication that these cancellations may persist at all loops in 4D $\cN\geq 5$, see more discussion on this in   \cite{Kallosh:2024ull} and in earlier studies in 
\cite{Kallosh:2010kk} and in this book in \cite{Nicolai:2024hqh}. The existence of ``enhanced dualities" is established in \cite{Kallosh:2024ull} for any loop order for 4D $\cN\geq 5$ supergravity theories. It is supported by the available computation in 4D $\cN=5$ at 4 loops, but the prediction is valid for any loop order for 4D $\cN\geq 5$ supergravity theories.

\section{  Attractors in cosmology }
\subsection{Dark energy and de Sitter supergravity}\label{Sec:dS}
Since the discovery of dark energy almost three decades ago, one of the simplest explanations of dark energy is via a positive cosmological constant $\Lambda$. It was possible to find de Sitter vacua in $\cN=1$ matter-coupled supergravity. For example, in the Polonyi model with one chiral superfield, there is a choice of a parameter $|\beta| < 0.268$, which leads to a minimum at positive $V$ and to asymptotically de Sitter universe \cite{Kallosh:2002gf}.
But it remained difficult to explain the tiny value of the cosmological constant  $\sim 10^{-120} M_P^4$. 

Over the years, there has been an increasing amount of observational evidence for an accelerating universe, where a positive cosmological constant is a good fit for the data.  All efforts were made to understand it better in the context of string theory and supergravity. In string theory, the KKLT construction was proposed \cite{Kachru:2003aw}, including an anti-D3 brane uplifting mechanism associated with the Volkov-Akulov model \cite{Volkov:1973ix}.

A component supergravity action with spontaneously broken local supersymmetry, generalizing the globally supersymmetric Volkov-Akulov model \cite{Volkov:1973ix} was constructed in \cite{Bergshoeff:2015tra}. It describes supergravity interacting with a nilpotent multiplet. It has a de Sitter vacuum even in pure supergravity without matter multiplets.

In the past, the cosmological constant was known to be negative or zero in pure supergravity without scalar fields.  In \cite{Bergshoeff:2015tra}  supersymmetry is spontaneously broken and non-linearly realized, so there is no conflict with no-go theorems that prohibit positive $\Lambda$ with linearly realized supersymmetry. A complete action with non-linearly realized supersymmetry before gauge-fixing local supersymmetry is presented in  \cite{Bergshoeff:2015tra}. Here we will only show the action {\it in the unitary gauge where the spinor of the nilpotent multiplet vanishes}.
\be
e^{-1}{\cal L}|_{\psi_X=0} ={1\over 2} \Big[ R( e, \omega(e)) - \bar \psi_\mu \gamma^{\mu\nu\rho}D_\nu \psi_\rho + m_{3\over 2} \bar \psi_\mu \gamma^{\mu\nu} \psi_\nu +{\cal L}_{SG, torsion}\Big]  + {3  m^2_{3/ 2}}- F_X^2 \ .
\ee
Here $F_X$ is the non-vanishing value of the auxiliary field of the nilpotent multiplet,  Volkov-Akulov uplifting constant, and $m_{3/ 2}$ is a mass of gravitino.
The cosmological constant is a difference between two contributions, 
\be
\Lambda = F_X^2- {3  m^2_{3/ 2}} \ .
\ee
It allows the multiverse interpretation, where $\Lambda$, $F_X$, $m_{3/ 2}$ may take different values in different exponentially large parts of the universe created by inflation. But life as we know it is possible only if $\Lambda$ is tiny, as indicated by cosmological observations. Indeed,  galaxies would not form in the universe with a large positive $\Lambda$, and the universe with a large negative $\Lambda$ would rapidly collapse \cite{Linde:1984ir}.

\
\subsection{CMB data and inflationary $\a$-attractor models}\label{Sec:alpha}
For supergravity experts we will present here the basic definition of inflationary observables and how these are related to theoretical inflationary models we build. In Fig. \ref{fig:LiteBIRD}, the predictions for CMB observables $r$ versus $n_s$ are superimposed with the data from various current and future experiments.
Primordial power spectra are conventionally parameterized as
\be
\Delta^2_{\zeta} (k)= \Delta^2_{\zeta} \Big ({k\over k_*}\Big )^{n_s(k)-1} \ .
\ee
The ratio of the power in primordial gravitational waves to the power in primordial density perturbations:
tensor-to-scalar ratio $r $ is
\be
r= {\Delta^2_{h} (k)\over \Delta^2_{\zeta} (k)} \ .
\ee 
The values of $n_s, r$ can be calculated for any inflationary model and compared with the data. The observational bound on $r$  is currently   $r< 0.028$ \cite{Galloni:2022mok}. 

The light blue areas shown in Fig. \ref{fig:LiteBIRD} in the Introduction represent the values of $n_s$ and $r$ consistent with BICEP/Keck (BK) and Planck data. The green/yellow area is where one hopes to get the data from LiteBIRD.  One can see that inflationary models with simple monomial potentials like  $V\sim \phi^n$ (a dark blue region in Fig. \ref{fig:LiteBIRD}) are in tension with the data at more than $2\sigma$ level.

The general 4D $\cN=1$ supergravity interacting with chiral multiplets requires the scalar manifold to be a \K\, manifold. In particular,  the \K\, manifold for one chiral multiplet with the  coset space ${SL(2,\mathbb{R})\over U(1)}$ defines the $\a$-attractor models \cite{Kallosh:2013yoa} with any value of $\a$. This choice is motivated by extended supergravities with $\cN\geq 2$. 
However, if the choice is motivated by string theory or by compactification from higher dimensions, a restriction is that $3\a=n$ is an integer,  $1\leq n\leq 7$ \cite{Ferrara:1989bc,Ferrara:2016fwe}.
  
In half-plane variables with ${\rm Re} \, T>0$
\be
K(T, \bar T)= -3\a \log(T+\bar T) \qquad \Rightarrow  \qquad e^{-1}{\cL}_{kin}=- 3\a {g^{\mu\nu} \partial_\mu T \partial_\nu \bar T \over (T+\bar T)^2} \ .
\label{K} \ee
  Here the parameter $\a$ is related to a \K\, curvature, as  was first observed in \cite{Ferrara:2013rsa} in the context of inflation in supergravity
\be
R_K= -{2\over 3\a} \ .
\ee
Using Cayley transform $T={1+Z\over 1-Z}$ one can switch to disk variables $|Z\bar Z<1$ where
\be
e^{-1}{\cL}_{kin}=- 3\a {g^{\mu\nu} \partial_\mu Z \partial_\nu \bar Z \over (1- Z \bar Z)^2}  \ .
\label{disk}\ee

The simplest $\a$-attractor model  just adds to a kinetic term in disk variables in eq.\rf{disk} a potential $|Z|^{2n}$ so that the total scalar part of the inflationary model is
\be
e^{-1}{\cL} ={1\over 2} R -3\a {g^{\mu\nu} \partial_\mu Z \partial_\nu \bar Z \over (1- Z \bar Z)^2 }- V_0 \, |Z|^{2n} \ .
\label{disk1}\ee
The single scalar T-model $\a$-attractors describe the models where the field $Z-\bar Z$ (the axion) is stabilized and vanishes during inflation, and inflation is driven by the inflaton field $\vp$ such that 
$Z=\bar Z= \tanh (\vp/\sqrt{6\a})
$.\footnote{Stabilization of the inflaton partner is achieved in the supergravity theory with an additional nilpotent chiral multiplet \cite{Carrasco:2015rva,Kallosh:2017wnt}. We also refer the reader to a contribution to this book on non-linear supergravity and inflationary cosmology, including the nilpotent multiplet \cite{Antoniadis:2024hvw}.}.   
The model with stabilized axion is
\be
{1\over 2} R -{1\over 2} (\partial\vp)^2 - V_0 \, \tanh^{2n} (\vp/\sqrt{6\a}) \ .
\label{Tmodel}\ee
\begin{figure}[t]
\centering
\includegraphics[width=125mm]{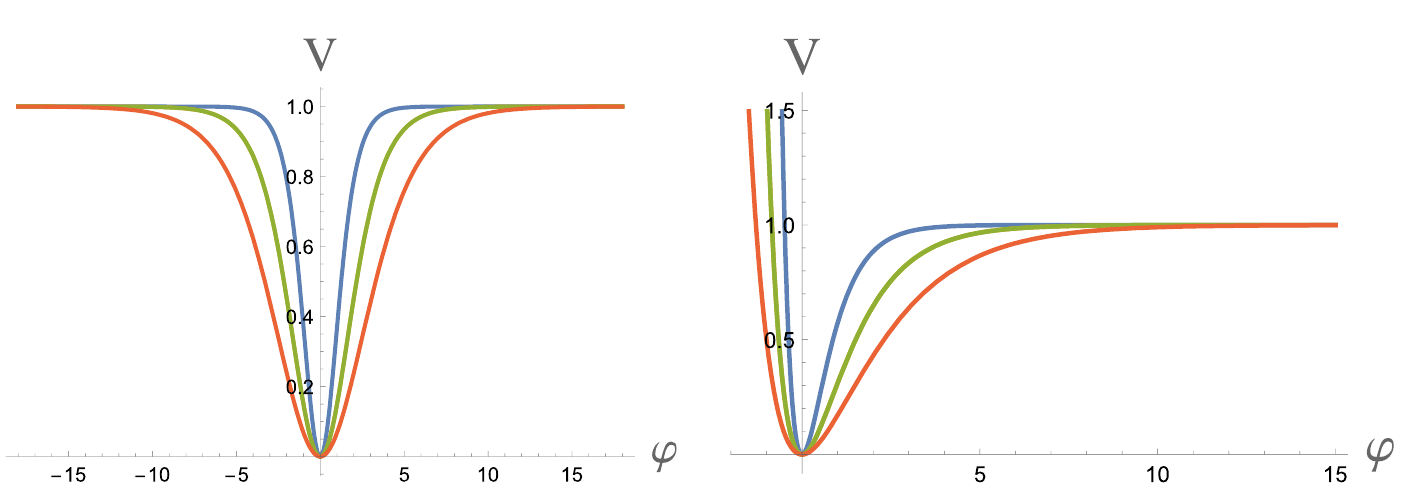}
\caption{\small On the left there is  T-model defined in \rf{Tmodel} for $n=1$, $V_0=1$, $3\a=1,3,7$. On the right, there is an E-model potential in \rf{Emodel} for the same parameters. The E-model potential with  $\a=1$ coincides with the potential in the Starobinsky model.}
\label{fig:TT}
\end{figure}
 In the slow roll approximation, the cosmological observables are given by the following expressions, which are also plotted in the left part of Fig. \ref{fig:CosmAttractors}:
\begin{align}
n_s(\alpha, n, N)  = {1 - {2 \over N}   - {3 \alpha \over 4 N^2} +{ 1\over 2nN}(1 - {1\over N})g(\alpha,n)
\over 1+ {1\over 2nN} g(\alpha,n) + {3 \alpha\over 4N^2}} \, ,  \quad 
r (\alpha, n, N)  = {12  \alpha\over N^2+ {N\over 2n} g(\alpha,n) + {3\over 4}\alpha} \, , \label{exact}
\end{align}
where
\be
g(\alpha,n)\equiv \sqrt{3 \alpha (4 n^2 + 3 \alpha)} \ .
\ee
Here $N$ is the number of e-folding of inflation, which is $N\sim 55$.
 In the  limit $\alpha \to \infty$ one has 
 \begin{align}
  n_s & = 1-\frac{2n+2}{2N+n}\,, \qquad 
  r = \frac{16 n}{2N+n} \, ,
 \end{align}
 which coincide with the corresponding expressions for the theory $V(\vp) \sim \vp^{2n}$. We therefore recover from all models of the type $\tanh^{2n} (\varphi/\sqrt{6\alpha})$ the corresponding  monomial models $\vp^{2n}$. 
 
We now look at the small $\alpha$ behavior close to the attractor, where $\alpha$ is of order one. Expanding \eqref{exact} in the large-$N$ limit with $N\approx 55$ we find
 \begin{align}\label{attractor2}
  n_s & \approx 1-\frac{2}{N}+\frac{\sqrt{3 \alpha (4 n^2 + 3 \alpha)} - 3 n \alpha}{2 n N^2} \,, \qquad
  r \approx \frac{12 \alpha}{N^2}-\frac{6 \alpha \sqrt{3 \alpha (4 n^2 + 3 \alpha)}}{n N^3} \,.
 \end{align}
The attractor point where all $V=\tanh^{2n} (\varphi/\sqrt{6\alpha})$ models tend to  universal values of $n_s, r$  is close to $\a\sim 1$ 
 and below where the last terms in equations defining both $n_s$ and $r$ can be neglected and we find $n$-independent values
 \begin{align}\label{attractor3}
  n_s & \approx 1-\frac{2}{N} \,, \qquad
  r \approx \frac{12 \alpha}{N^2} \,.
 \end{align} 
 In Fig. \ref{fig:LiteBIRD} on the left, one can see a grey band describing $V=\tanh^{2} (\varphi/\sqrt{6\alpha})$ model prediction for cosmological observables. The CMB experiments agree with the values of $n_s$ for all $\a$-attractors. The latest values are given in  \cite{SPT-3G:2024atg} and  \cite{Galloni:2022mok}.
However, primordial gravitational waves have not been discovered yet; there is only a bound $r< 0.028$  \cite{Galloni:2022mok}. If future experiments detect primordial gravitational waves, i.e. if the actual value of $r$ is known, in the context of $\a$-attractors, these measurements will tell us the value of the \K\, curvature of the field space 
$
R_K= -{2\over 3\a} 
$.

The E-models of $\a$-attractors are more natural in half-plane variables \rf{K}. They take the form
\be
e^{-1}{\cL} ={1\over 2} R -3\a {g^{\mu\nu} \partial_\mu T \partial_\nu \bar T \over (T+\bar T)^2 }- V_0 \, (T-1)^{2n} \ .
\label{E}\ee
The single scalar E-model $\a$-attractors is a case 
$
T=\bar T= e^{-\sqrt{2/3\a} \, \vp}
$ and the field $T-\bar T=0$ is stabilized and we get
\be
{1\over 2} R -{1\over 2} (\partial\vp)^2 - V_0 \, (1-e^{-\sqrt{2/3\a} \, \vp})^ {2n}  \ .
\label{Emodel}\ee
Now consider {\it discrete targets} in the right panel of Fig.  \ref{fig:LiteBIRD} known as seven Poincare disks. These are models shown in eq. \rf{disk1} where 
\be
3\a=7,6,5,4,3,2,1
\ee
 These models with discrete values of $3\a$ were proposed and studied in \cite{Ferrara:2016fwe}. They originate from 
 compactification of 11D M theory on a 7 manifold with $G_2$ holonomy to 4D minimal supergravity or from 
 compactification of 10D superstring theory on a 6-torus to 4D minimal supergravity.
According to the relation $r \approx \frac{12 \alpha}{N^2} $,  the values of $r$ for each of these models, for $N\approx 55$ are
\be
r\approx \{9.1, \, 7.8, \, 6.5 , \, 5.2, \, 3.9, \, 2.6, \, 1.3\}\times 10^{-3} \, .
\ee
In 2015, BICEP2/Keck Array and Planck had a bound $r< 0.11$; in 2022, it was $r<0.028$. The $3\alpha = 7$ disk model has $r\approx 0.0091$. Hopefully, this level will be reached relatively soon. The next one, $3\alpha = 6$, is quite interesting: in addition to representing the second from the top Poincar\'e disk model of $\alpha$-attractors with $\alpha = 2$, the value $r\approx 0.0078$ is also a value predicted by the Fibre inflation model in string theory, see the most recent paper \cite{Cicoli:2024bxw}  and references therein. It is known \cite{Kallosh:2017wku} that Fibre inflation can be effectively described as a supergravity $\alpha$-attractor with $\alpha = 2$ . Thus, the level at $r\sim 0.0078$ is degenerate. The case $\alpha =1$ is also degenerate; it appears in the Starobinsky model, in the Higgs inflation model, in the superconformal attractor model \cite{Kallosh:2013hoa}, as well as in the $\alpha$-attractors with $\alpha = 1$ \cite{Kallosh:2013yoa}.

The smallest Poincar\'e disk $3\a=1$ is especially interesting: it is the last discrete $\alpha$-attractor target associated with string theory, and compactification from higher dimensions, $r\sim 0.0013$. In the absence of a detection, LiteBIRD will set an upper limit of $r < 0.002$ at 95 \% C.L. But this will not exclude the models with $3\a=1$  and $r\sim 0.0013$. 
If primordial gravitational waves are not detected at the level $r\sim 10^{-3}$, it will still leave us with generic $\cN=1$ supergravity targets where $\a$ is not constrained, as we see in the grey band at the left panel in Fig. \ref{fig:LiteBIRD}.

\subsection{$SL(2,\mathbb{Z})$   inflation }\label{nonlinsuper}

The new supergravity class of inflationary models  with $SL(2,\mathbb{Z})$  symmetry  in \cite{Kallosh:2024ymt,Kallosh:2024pat,Kallosh:2024whb}
 promotes  the concept of a ``target space modular invariance'' \cite{Ferrara:1989bc}. In these models the \K\,  potential for a complex scalar field $T=-i\tau$ is $3\a \log (T+\bar T)$, where $3\a$ is an integer. 
 Target space duality different from duality symmetry in string theory since we study physics in 4D space-time, not on the 2D world-sheet. Here we refer the reader to a contribution to this book on 
 string dualities and modular symmetries in supergravity \cite{Cribiori:2024qsv}.

The reason why, in the past, it was difficult to construct modular inflation models in supergravity is that the  \K\, invariant supergravity function
$
\mathcal G(T, \bar T)
$ dependent only of modulus $T$ without the nilpotent multiplet.
 The advantage of the nilpotent superfield for constructing modular invariant cosmological models will be now demonstrated.

In the framework of $\overline {D3}$ induced geometric inflation   \cite{Kallosh:2017wnt} supergravity is defined by a function $\mathcal G = K+\ln |W|^2$ which includes in addition to our single superfield 
$
T= -i \tau = e^{{\sqrt {2\over 3\alpha}\vp}}-i \theta
$
 also a superfield $X$, which is nilpotent, i.e. $X^2=0$. It is a supergravity version of the uplifting $\overline {D3}$ brane, which supports de Sitter vacuum in supergravity\cite{Bergshoeff:2015tra}.

For a half-plane variable $T$, we consider the following  K\"ahler invariant function $\mathcal G$ 
\begin{align}
\mathcal G=&-3\alpha\ln(T+\bar T)+G_{X\bar X}(T, \bar T) X\bar X +\ln |W_0+F_X \, X|^2 \ .
\label{G}\end{align}
Here $X$ is a nilpotent superfield, $W_0$ is a constant defining the mass of gravitino, and $F_X$ is a constant, defining the auxiliary field vev.  In \cite{Kallosh:2024ymt} we made a  choice, following \cite{Kallosh:2022vha}
\begin{align}
G_{X\bar X}(T, \bar T)=\frac{|F_X|^2}{(T+\bar T)^{3\alpha}[ \Lambda  +V(T,\bar T)]+3|W_0|^2(1-\alpha)}\, , \qquad \Lambda =F_X^2 - 3W_0^2 \ .
\label{GXX}\end{align}
where $V(T,\bar T)$ are $SL(2,\mathbb{Z})$ invariant potentials with Minkowski minima presented in \cite{Kallosh:2024ymt,Kallosh:2024pat,Kallosh:2024whb}.
In this case, the bosonic action following from this supersymmetric construction is 
\be
{ {\cal L} (T, \bar T)\over \sqrt{-g}} =  {R\over 2} - {3\alpha\over 4} \, {\partial T \partial \bar T\over ({\rm Re} \,  T )^2}- [\Lambda +V(T, \bar T)]  \ .
\label{hyper2}\ee
If $\Lambda =|F_X|^2 -3|W_0|^2>0$  and at the end of inflation $V(T, \bar T)=0$ there is an exit into a de Sitter vacuum. This action is 
$SL(2,\mathbb{Z})$ invariant if $V(T, \bar T)$ is $SL(2,\mathbb{Z})$ invariant.

Consider supergravity construction defined in eqs. \rf{G},  \rf{GXX}. It appears that $G_{X\bar X}< 0$ at the minimum of   $V(T, \bar T)$  if $\Lambda\ll W_0$ and $\a>1$.  Is it a problem for modes with  $\a>1$?

Consider the full supergravity action where $G_{X\bar X}$ is present and might be affected by $G_{X\bar X}$ being negative. These are  kinetic terms in the action of the form
\be
G_{X\bar X} (\partial_\mu X \partial^\mu \bar X + \bar \psi^X \gamma_\mu D^\mu \psi^X)
\label{kin}\ee
If $X$ would be a normal chiral multiplet one would have to require that $G_{X\bar X}> 0$. However, the nilpotent multiplet in supergravity satisfies the constraint \cite{Bergshoeff:2015tra} that the scalar  depends on the square of the spinor field 
\be
X= {(\psi^X)^2\over 2 F^X}
\label{X} \ee
where $F^X$ is the auxiliary field of the nilpotent multiplet.
This makes the 1st term in eq. \rf{kin} quartic in spinor field which does nor require $G_{X\bar X}$ to be positive anymore. The second term in eq. \rf{kin}  is quadratic in spinors and it presence in the action  might raise the issue of consistency of supergravity with  negative $G_{X\bar X}$.

Fortunately,  we can use local supersymmetry to gauge fix the fermion from the nilpotent multiplet to vanish, $\psi^X=0$. This is a unitary gauge discussed in \cite{Bergshoeff:2015tra} in supergravity interacting with one nilpotent multiplet and in \cite{Kallosh:2015sea} in supergravity interacting with a nilpotent multiplet and other chiral multiplets. The scalar field of the nilpotent multiplet depends on its fermion as shown in eq. \rf{X}.  In the unitary gauge $\psi^X=0$,   the sign of $G_{X\bar X}(T, \bar T)$ does not matter since kinetic terms of the boson $X$ field and of the fermion $\psi_X$ field are both absent. Alternatively, one can take a unitary gauge $v=0$ where goldstino is
$
v= {1\over \sqrt 2} e^{K(T, \bar T) /2} (\psi^T D_T W + \psi^X D_X W) +{1\over 2}iP_L \lambda^A {\cal P}_A $.
In the gauge where $v=0$, the fermion $\psi^X$ is replaced by a nonlinear function of moduli $T$ and a fermion  $\psi^T$. Therefore, the kinetic terms of the boson $X$ and the fermion $\psi^X$ with a negative $G_{X\bar X}$ are not harmful. 
Thus {\it supergravity defined by $\mathcal G$ in eqs. \rf{G}, \rf{GXX}  is consistent for any $\a$} as we have shown here using unitary gauges for local supersymmetry.

It is interesting to compare our results here with the ones that can be obtained in the context of liberated supergravity\cite{Farakos:2018sgq} \footnote{In \cite{Jang:2020cbe} liberated $\cN = 1$ supergravity was used as EFT for describing  inflationary dynamics.}, where in the case of one matter multiplet and a nilpotent multiplet 
\be
G_{X\bar X}= {e^K(T, \bar T) \over {\cal U} (T, \bar T)} \ .
\ee
The proposal in \cite{Farakos:2018sgq} is that the Green function of the nilpotent field involving ${\cal U} (T, \bar T)$ is an arbitrary function and does not have to be positive. This is valid under the condition that $\langle D_T W\rangle \neq 0$, so that it is possible to use a gauge-fixing condition $v=0$ eliminating goldstino.

Meanwhile, in \cite{Kallosh:2015sea}, the unitary gauge $v=0$ was compared with the gauge $\psi^X=0$. In case  $v=0$ gravitino is not mixed with other fermions of the theory, in gauge $\psi^X=0$, the Lagrangian is simplified significantly. In any case, we find full agreement between our analysis of supergravity with $SL(2,\mathbb{Z})$  cosmology in eqs. \rf{G}, \rf{GXX} and the setup  in  \cite{Farakos:2018sgq}.  Namely, consistency of supergravity does not impose any restrictions on \K\, curvature $R_K= -{2\over 3\a}$.

This means that using the methods described here, one can embed any bosonic $SL(2,\mathbb{Z})$ invariant model into supergravity with one chiral superfield  $T=-i\tau$ and a nilpotent superfield $X$.

In eq. \rf{hyper2}  we provided a bosonic part of the action of the supergravity defined in eqs. \rf{G},  \rf{GXX} which has a de Sitter exit from inflation with $\Lambda =|F_X|^2 -3|W_0|^2>0$. This gives us a supersymmetric generalization of the bosonic theories which we constructed and applied to cosmology. 
   
Consider one of the simplest potentials of this type,
\be\label{Renata}
V(\tau, \bar \tau)= V_0 \Big (1-{\ln |j^2(i)|
\over \ln ( | j(\tau))|^2 +  j^{2}(i))}  \Big ) \ , \qquad \tau=iT \ .
\ee
Here $j(i)= 12^3$ corresponds to Absolute Klein invariant $J(\tau)= 12^{-3} j(\tau)$ taking a value $J(i)=1$.

\begin{figure}[t]
\centering
\includegraphics[width=70mm]{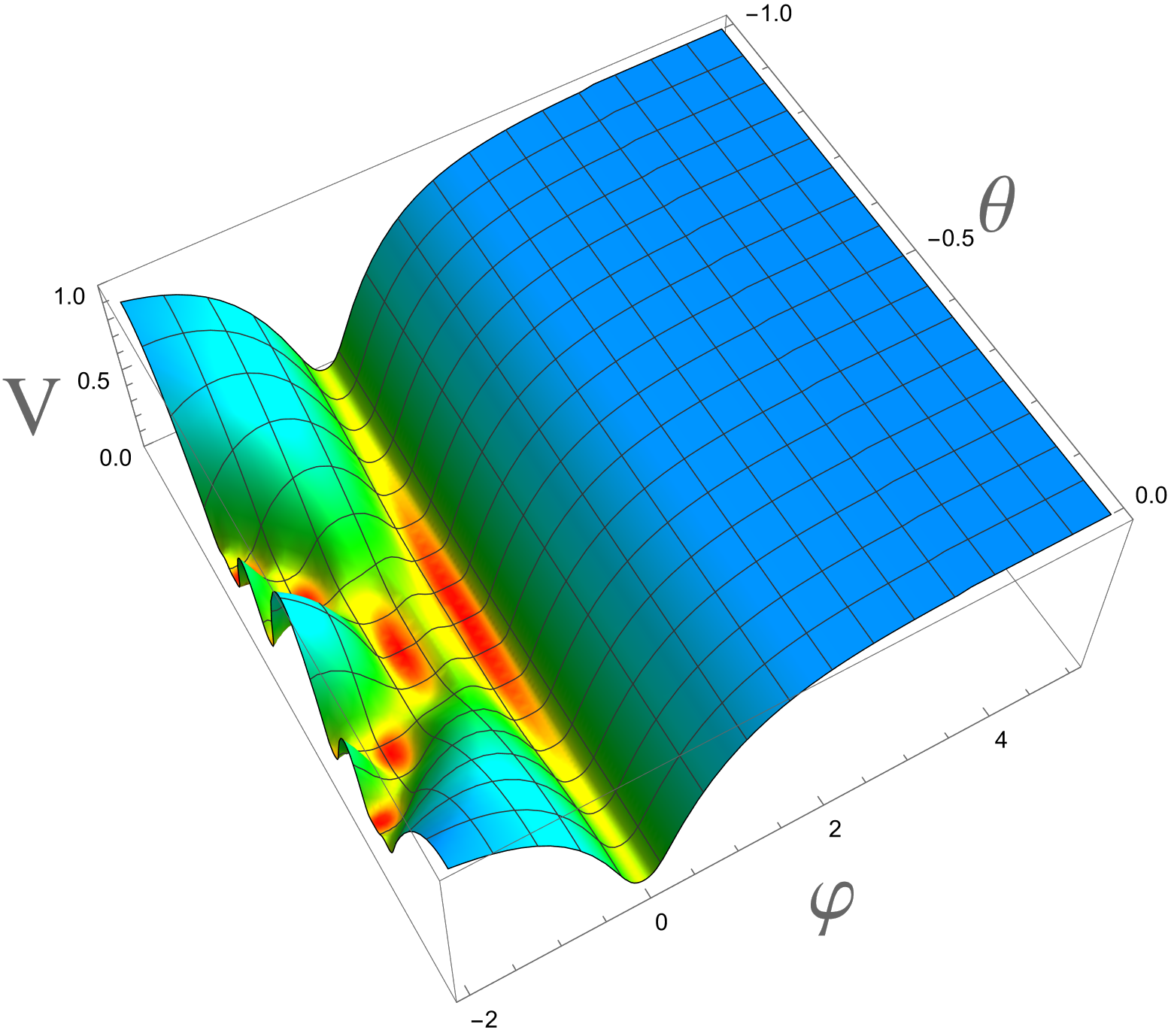}
\caption{\small Potential \rf{Renata}  for  $V_{0} = 1$, $  \tau = \theta+ i e^{{\sqrt {2\over 3\alpha}}} \theta$, $\alpha=1/3$. The potential is non-negative. The height of the potential is color-coded, from blue to red. The blue plateau approaches $V_{0}= 1$ at $\vp \to +\infty$. Red spots correspond to $V \ll 1$, which helps to visually identify the minima of the potential. All minima have the same depth $V = 0$, but one can uplift all of them by adding a tiny constant $\Lambda$ to the potential. This is consistent with the bosonic version \rf{hyper2} of the supergravity defined by eqs. \rf{G}, \rf{GXX}. }
\label{fig:f2}
\end{figure}
The basic difference with $\a$-attractors described in  Sec. \ref{Sec:alpha} is that the potentials in $SL(2,\mathbb{Z})$   models preserve the discrete subgroup of $SL(2,\mathbb{R})$ group,  which is a symmetry of the kinetic term.  The $SL(2,\mathbb{Z})$   invariant potentials are more complicated than  simple $\a$-attractors since they depend on modular invariants like 
\be
j(\tau) = q^{-1} + \sum_{n=0} c_n q^n\, , \qquad q= e^{2\pi i\tau}  \ .
\label{series}
\ee
These potentials  during inflation 
have plateau potentials with respect to the inflaton and axion fields. The slope of the potential in the inflaton direction is exponentially suppressed, but  the slope of the potential in the axion direction is double-exponentially suppressed \cite{Kallosh:2024whb}.

 We have recently introduced these models with axion field stabilized  \cite{CKLR}, to avoid generation of isocurvature perturbations. 
\be\label{Renata}
V_{stab} (\tau, \bar \tau)= V_0 \Big (1-{\ln |j^2(i)|
\over \ln ( | j(\tau))|^2+A\,
| j(\tau) \pm\overline {j(\tau)}|^{2} +  j^{2}(i))}  \Big ).
\ee
  We have found in \cite{CKLR} that these models with stabilized axions have the same features as $\a$-attractors described in  Sec. \ref{Sec:alpha}. Namely, they give universal predictions for inflationary observables like the ones in eq.\rf{attractor3}.

We have investigated in \cite{Kallosh:2024pat} the global structure of the recently discovered family of $SL(2,\mathbb{Z})$-invariant potentials describing inflationary $\alpha$-attractors in \cite{Kallosh:2024ymt}. These potentials have an inflationary plateau consisting of the fundamental domain and its images fully covering the upper part of the Poincar\'e half-plane. Meanwhile, the lower part of the half-plane is covered by an infinitely large number of ridges, which, at first glance, are too sharp to support inflation. However, one can show that this apparent sharpness is just an illusion created by hyperbolic geometry, and each of these ridges is physically equivalent to the inflationary plateau in the upper part of the Poincar\'e half-plane.

The way to see it is to switch from the half-plane axion-inflaton $\tau = x+iy= \theta+i e^{\sqrt{2\over 3\alpha}  \vp } $ coordinates 
with a kinetic term 
\begin{equation}\label{kin1}
{\cal L}_{kin}^{axion-inflaton}=\frac12(\partial\vp)^2+\frac{3\alpha}{4}e^{-2\sqrt{\frac{2}{3\alpha}}\vp}(\partial\theta)^2 
\end{equation}
to Killing coordinates $\tau = i e^{\sqrt{2\over 3\alpha} (\tilde\vp -i\vartheta)} $ used for the investigation of $\alpha$-attractors in   \cite{Carrasco:2015rva}  with the kinetic term 
\be
{\cal L}_{kin}^{Killing}={1\over 2}{ (\partial \tilde\vp)^2+ (\partial\vartheta)^2\over \cos^2 (\sqrt{2\over 3\alpha} \vartheta)} \ .
\label{Lkin}\ee

\begin{figure}[h!]
\centering
\includegraphics[scale=0.46]{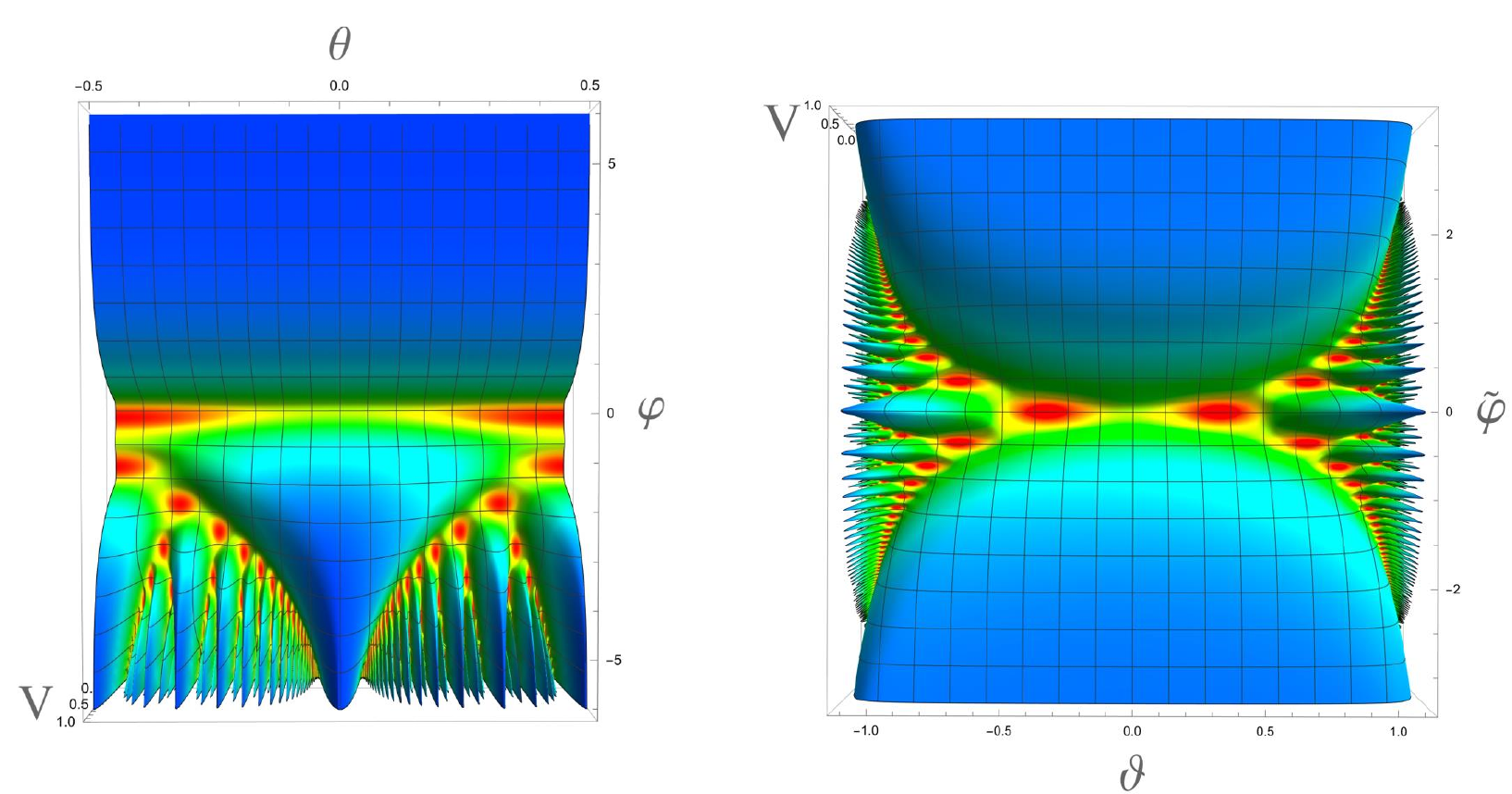}
\vskip -10pt
\caption{\footnotesize  Top view on the potential \rf{Renata}: on the left  as a function of $\theta$ and $\vp$ for  $3\alpha=2$. There is an inflationary plateau at  $\vp > 0$, the minima at $\theta  =  -0.5,  0.5$, and a saddle point at $\theta = 0$. At  $\vp < 0$, one can see a complicated profile of multiple mountains and a proliferation of minima and saddle points. Same potential \rf{Renata} in the right panel as a function of $\tilde{ \vp}, \vartheta $. In Killing coordinates the potential has a symmetry $\tilde \vp \to -\tilde \vp$ and $\vartheta \to -\vartheta$}
\label{SL2z}
\end{figure}

The potential in the right panel in Fig. \ref{SL2z} in Killing coordinates in the region at small  $\vartheta$ has both coordinates with canonical kinetic terms since $ \cos^2 (\sqrt{2\over 3\alpha} \vartheta) \to 1$ in \rf{Lkin}. Meanwhile, in half-plane axion-inlaton variables in the left panel in Fig. \ref{SL2z} , the distance between points with coordinates $\theta$ is $\sim e^{-2\sqrt{\frac{2}{3\alpha}}\vp}$. The physical distance is  $e^{-\sqrt{\frac{2}{3\alpha}}\vp} d\theta$. At large negative $\vp$ this distance $\to \infty$. This is why we see sharp ridges at large negative $\vp$ in the left panel in Fig. \ref{SL2z}. Instead, in Killing coordinates in the right panel, these ridges are stretched and become plateaus. We show an example of one of the ridges stretched to a plateau in Fig. \ref{SL2z}. In fact, one can show the same for all ridges using $SL(2,\mathbb{R})$ transformations, which keep the kinetic term invariant but change the shape of the potential so that each ridge stretches into a plateau.

The fascinating part of this story is that in these theories with target space modular invariance, the global structure of the $SL(2,\mathbb{Z})$ invariant potentials involves an infinite number of saddle points, minima, and plateaus.

From the point of view of the fundamental domain, these are the $SL(2,\mathbb{Z})$  images of the plateau, saddle point, and a minimum of the potential with values inside or on the fundamental domain boundary. However, in cosmological applications, we have to take into account that only the total hyperbolic half-plane is geodesically complete.  We are running classical inflaton-axion trajectories during inflation superimposed over the potentials \cite{CKLR}. We find that the field approaches the minimum of the potential, which is at the boundary of the fundamental domain with $\tau \bar \tau =1$. It oscillates near the minimum crossing the boundary. We have also found trajectories starting from the fundamental domain plateau and not stopping at the nearest minimum at the boundary of the fundamental domain, but reaching out to other  minima at $\tau \bar \tau <1$.
Thus, the landscape of $SL(2, \mathbb{Z})$ invariant cosmological models has very interesting properties.

\section{Concluding remarks}
The studies of supergravity attractors in black hole physics and cosmology started almost three decades ago. Fortunately, the  ``Supergravity'' book \cite{Freedman:2012zz} and various contributions to this book \cite{Ortin:2024slu,Nicolai:2024hqh,Antoniadis:2024hvw,Cribiori:2024qsv,Cassani:2025sim} cover relevant progress in related areas of supergravity.
Therefore, we focused on earlier investigations of attractors only to the extent relevant to important recent supergravity developments, both in black holes and in cosmology. Here is the summary of the important issues discussed in this article: 
\begin{itemize}
  \item  1/8 BPS and Non-BPS extremal black holes in maximal $\cN=8$ supergravity have a natural embedding into a standard 4D supergravity
  \cite{Cremmer:1979up}  and a non-standard one \cite{Andrianopoli:2002mf} reduced from 5D without dualization, respectively. Recognition of the difference between these two supergravities of type I and type II \cite{Kallosh:2024rdr} has led to the concept of {\it enhanced dualities} in  \cite{Kallosh:2024ull}.
  Enhanced dualities explain the mysterious {\it enhanced cancellation of ultraviolet divergences} in 82 Feynman diagrams in 4-loop superamplitude in $\cN=5$ supergravity  \cite{Bern:2014sna, Bern:2023zkg}. Enhanced dualities may have important implications for the possibility of the all-loop finiteness of 4D $\cN >4$ supergravities discussed in \cite{Kallosh:2024ull,Kallosh:2010kk} and in this book in \cite{Nicolai:2024hqh}.

    \item We described supergravity realization of inflationary $\a$-attractors \cite{Kallosh:2013yoa}, where the moduli space has $SL(2, \mathbb{R})$ symmetry and the \K\, curvature  is equal to $R_K=-{2\over 3\a}$. These models predict inflationary observables which are in agreement with available CMB data. They also predict the level of primordial gravitational waves, depending on \K\, curvature,  which will be tested in future CMB experiments, such as LiteBIRD \cite{LiteBIRD:2022cnt}, see Fig. \ref{fig:LiteBIRD}, and earlier by  BICEP/Keck, Simons observatory and other ground based experiments.
  \item We have presented recent supergravity versions of $SL(2, \mathbb{Z})$ invariant cosmological models \cite{Kallosh:2024ymt,Kallosh:2024pat,Kallosh:2024whb}.  We have shown in \cite{CKLR} that when axion is stabilized these models  have predictions analogous to $\a$-attractors and therefore compatible with the CMB data. These models have many interesting features to be studied in the future.  
\end{itemize}

\section*{\sc Acknowledgements}
We thank A.~Ceresole and G.~Dall'Agata for the kind invitation to contribute to this volume.
 We are grateful to  E.A. Bergshoeff,  J.J. Carrasco, S. Ferrara,  D.Z. Freedman, M. Gunaydin, H. Nicolai,
D. Roest,  R. Roiban, H. Samtleben, T. Ortin,  A. Van Proeyen, T. Wrase, and Y. Yamada,  our collaborators on the projects relevant to this article. We had useful discussions with  E. Copeland, D. Lust, A. Maharana and F. Quevedo at Strings 2025.
This work is supported by SITP and US National Science Foundation grant PHY-2310429.


\begin{thebibliography}{99}

\bibitem{Kallosh:2013yoa}
R.~Kallosh, A.~Linde and D.~Roest,
 ``Superconformal Inflationary $\alpha$-Attractors,''
JHEP \textbf{11}, 198 (2013)
doi:10.1007/JHEP11(2013)198
[arXiv:1311.0472 [hep-th]].
J.~J.~M.~Carrasco, R.~Kallosh, A.~Linde and D.~Roest,
 ``Hyperbolic geometry of cosmological attractors,''
Phys. Rev. D \textbf{92}, no.4, 041301 (2015)
doi:10.1103/PhysRevD.92.041301
[arXiv:1504.05557 [hep-th]].

\bibitem{Planck:2018jri}
Y.~Akrami \textit{et al.} [Planck],
 ``Planck 2018 results. X. Constraints on inflation,''
Astron. Astrophys. \textbf{641}, A10 (2020)
doi:10.1051/0004-6361/201833887
[arXiv:1807.06211 [astro-ph.CO]].

\bibitem{BICEP:2021xfz}
P.~A.~R.~Ade \textit{et al.} [BICEP and Keck],
 ``Improved Constraints on Primordial Gravitational Waves using Planck, WMAP, and BICEP/Keck Observations through the 2018 Observing Season,''
Phys. Rev. Lett. \textbf{127}, no.15, 151301 (2021)
doi:10.1103/PhysRevLett.127.151301
[arXiv:2110.00483 [astro-ph.CO]].

\bibitem{LiteBIRD:2022cnt}
E.~Allys \textit{et al.} [LiteBIRD],
 ``Probing Cosmic Inflation with the LiteBIRD Cosmic Microwave Background Polarization Survey,''
PTEP \textbf{2023}, no.4, 042F01 (2023)
doi:10.1093/ptep/ptac150
[arXiv:2202.02773 [astro-ph.IM]].



\bibitem{Ferrara:2016fwe}
S.~Ferrara and R.~Kallosh,
 ``Seven-disk manifold, $\alpha$-attractors, and $B$ modes,''
Phys. Rev. D \textbf{94}, no.12, 126015 (2016)
doi:10.1103/PhysRevD.94.126015
[arXiv:1610.04163 [hep-th]].
R.~Kallosh, A.~Linde, T.~Wrase and Y.~Yamada,
 ``Maximal Supersymmetry and B-Mode Targets,''
JHEP \textbf{04}, 144 (2017)
doi:10.1007/JHEP04(2017)144
[arXiv:1704.04829 [hep-th]].

\bibitem{Ortin:2024slu}
T.~Ort\'\i{}n,
``Black hole solutions in theories of supergravity,''
[arXiv:2412.12020 [hep-th]].


\bibitem{Ferrara:1995ih}
S.~Ferrara, R.~Kallosh and A.~Strominger,
 ``N=2 extremal black holes,''
Phys. Rev. D \textbf{52}, R5412-R5416 (1995)
doi:10.1103/PhysRevD.52.R5412
[arXiv:hep-th/9508072 [hep-th]].
S.~Ferrara and R.~Kallosh,
 ``Supersymmetry and attractors,''
Phys. Rev. D \textbf{54}, 1514-1524 (1996)
doi:10.1103/PhysRevD.54.1514
[arXiv:hep-th/9602136 [hep-th]].
S.~Ferrara and R.~Kallosh,
 ``Universality of supersymmetric attractors,''
Phys. Rev. D \textbf{54}, 1525-1534 (1996)
doi:10.1103/PhysRevD.54.1525
[arXiv:hep-th/9603090 [hep-th]].
S.~Ferrara, G.~W.~Gibbons and R.~Kallosh,
 ``Black holes and critical points in moduli space,''
Nucl. Phys. B \textbf{500}, 75-93 (1997)
doi:10.1016/S0550-3213(97)00324-6
[arXiv:hep-th/9702103 [hep-th]].

\bibitem{Ferrara:2006em}
S.~Ferrara and R.~Kallosh,
 ``On N=8 attractors,''
Phys. Rev. D \textbf{73}, 125005 (2006)
doi:10.1103/PhysRevD.73.125005
[arXiv:hep-th/0603247 [hep-th]].

\bibitem{Ceresole:2009jc}
A.~Ceresole, S.~Ferrara, A.~Gnecchi and A.~Marrani,
 ``More on N=8 Attractors,''
Phys. Rev. D \textbf{80}, 045020 (2009)
doi:10.1103/PhysRevD.80.045020
[arXiv:0904.4506 [hep-th]].

\bibitem{Kallosh:2024rdr}
R.~Kallosh, H.~Samtleben and A.~Van Proeyen,
 ``Gauge-fixing local H symmetry in supergravities,''
JHEP \textbf{12} (2024), 027
doi:10.1007/JHEP12(2024)027
[arXiv:2409.18950 [hep-th]].

\bibitem{Nicolai:2024hqh}
H.~Nicolai,
``N=8 Supergravity, and beyond,''
[arXiv:2409.18656 [hep-th]].

\bibitem{Cremmer:1997ct}
E.~Cremmer, B.~Julia, H.~Lu and C.~N.~Pope,
 ``Dualization of dualities. 1.,''
Nucl. Phys. B \textbf{523}, 73-144 (1998)
doi:10.1016/S0550-3213(98)00136-9
[arXiv:hep-th/9710119 [hep-th]].


 
\bibitem{Cremmer:1979up}
E.~Cremmer and B.~Julia,
 ``The SO(8) Supergravity,''
Nucl. Phys. B \textbf{159} (1979), 141-212
doi:10.1016/0550-3213(79)90331-6
B.~de Wit and H.~Nicolai,
 ``N=8 Supergravity,''
Nucl. Phys. B \textbf{208} (1982), 323
doi:10.1016/0550-3213(82)90120-1


\bibitem{Andrianopoli:2002mf}
L.~Andrianopoli, R.~D'Auria, S.~Ferrara and M.~A.~Lledo,
 ``Gauging of flat groups in four-dimensional supergravity,''
JHEP \textbf{07}, 010 (2002)
doi:10.1088/1126-6708/2002/07/010
[arXiv:hep-th/0203206 [hep-th]].

\bibitem{Sezgin:1981ac}
E.~Sezgin and P.~van Nieuwenhuizen,
 ``Renormalizability Properties of Spontaneously Broken $N=8$ Supergravity,''
Nucl. Phys. B \textbf{195}, 325-364 (1982)
doi:10.1016/0550-3213(82)90403-5



\bibitem{Kallosh:2024ull}
R.~Kallosh,
``Enhanced Duality in 4D Supergravity,''
[arXiv:2405.20275 [hep-th]].
R.~Kallosh,
``The role of Sp(2n) duality in quantum theory,''
[arXiv:2410.19216 [hep-th]].


\bibitem{Bern:2014sna}
Z.~Bern, S.~Davies and T.~Dennen,
 ``Enhanced ultraviolet cancellations in $\mathcal N=5$ supergravity at four loops,''
Phys. Rev. D \textbf{90}, no.10, 105011 (2014)
doi:10.1103/PhysRevD.90.105011
[arXiv:1409.3089 [hep-th]].
\bibitem{Bern:2023zkg}
Z.~Bern, J.~J.~M.~Carrasco, M.~Chiodaroli, H.~Johansson and R.~Roiban,
``Supergravity amplitudes, the double copy and ultraviolet behavior,''
[arXiv:2304.07392 [hep-th]].



\bibitem{Kallosh:2013hoa}
R.~Kallosh and A.~Linde,
 ``Universality Class in Conformal Inflation,''
JCAP \textbf{07}, 002 (2013)
doi:10.1088/1475-7516/2013/07/002
[arXiv:1306.5220 [hep-th]].

\bibitem{SPT-3G:2024atg}
F.~Ge \textit{et al.} [SPT-3G],
 ``Cosmology From CMB Lensing and Delensed EE Power Spectra Using 2019-2020 SPT-3G Polarization Data,''
[arXiv:2411.06000 [astro-ph.CO]].
\bibitem{Galloni:2022mok}
G.~Galloni, N.~Bartolo, S.~Matarrese, M.~Migliaccio, A.~Ricciardone and N.~Vittorio,
 ``Updated constraints on amplitude and tilt of the tensor primordial spectrum,''
JCAP \textbf{04}, 062 (2023)
doi:10.1088/1475-7516/2023/04/062
[arXiv:2208.00188 [astro-ph.CO]].

\bibitem{Kallosh:2024ymt}
R.~Kallosh and A.~Linde,
``$SL(2,\mathbb{Z})$ Cosmological Attractors,''
[arXiv:2408.05203 [hep-th]].

\bibitem{Kallosh:2024pat}
R.~Kallosh and A.~Linde,
``Landscape of Modular Cosmology,''
[arXiv:2411.07552 [hep-th]].

\bibitem{Kallosh:2024whb}
R.~Kallosh and A.~Linde,
``Double Exponents in $SL(2,\mathbb{Z})$ Cosmology,''
[arXiv:2412.19324 [hep-th]].


\bibitem{Casas:2024jbw}
G.~F.~Casas and L.~E.~Ib\'a\~nez,
``Modular Invariant Starobinsky Inflation and the Species Scale,''
[arXiv:2407.12081 [hep-th]].

\bibitem{Ferrara:1989bc}
S.~Ferrara, D.~Lust, A.~D.~Shapere and S.~Theisen,
 ``Modular Invariance in Supersymmetric Field Theories,''
Phys. Lett. B \textbf{225}, 363 (1989)
doi:10.1016/0370-2693(89)90583-2
M.~Cvetic, A.~Font, L.~E.~Ibanez, D.~Lust and F.~Quevedo,
 ``Target space duality, supersymmetry breaking and the stability of classical string vacua,''
Nucl. Phys. B \textbf{361}, 194-232 (1991)
doi:10.1016/0550-3213(91)90622-5


\bibitem{Cribiori:2024qsv}
N.~Cribiori and D.~Lust,
``String dualities and modular symmetries in supergravity: a review,''
[arXiv:2411.06516 [hep-th]].

\bibitem{Kachru:2003aw}
S.~Kachru, R.~Kallosh, A.~D.~Linde and S.~P.~Trivedi,
``De Sitter vacua in string theory,''
Phys. Rev. D \textbf{68}, 046005 (2003)
doi:10.1103/PhysRevD.68.046005
[arXiv:hep-th/0301240 [hep-th]].

\bibitem{Antoniadis:2024hvw}
I.~Antoniadis, E.~Dudas, F.~Farakos and A.~Sagnotti,
``Non-Linear Supergravity and Inflationary Cosmology,''
[arXiv:2409.14943 [hep-th]].

\bibitem{Kallosh:2017wnt}
R.~Kallosh, A.~Linde, D.~Roest and Y.~Yamada,
 ``$ \overline{D3} $ induced geometric inflation,''
JHEP \textbf{07}, 057 (2017)
doi:10.1007/JHEP07(2017)057
[arXiv:1705.09247 [hep-th]].

\bibitem{Kallosh:2022vha}
R.~Kallosh and A.~Linde,
``Dilaton-axion inflation with PBHs and GWs,''
JCAP \textbf{08}, no.08, 037 (2022)
doi:10.1088/1475-7516/2022/08/037
[arXiv:2203.10437 [hep-th]].
Y.~Yamada,
``U(1) symmetric $\alpha$-attractors,''
JHEP \textbf{04}, 006 (2018)
doi:10.1007/JHEP04(2018)006
[arXiv:1802.04848 [hep-th]].
A.~Ach\'ucarro, R.~Kallosh, A.~Linde, D.~G.~Wang and Y.~Welling,
``Universality of multi-field $\alpha$-attractors,''
JCAP \textbf{04} (2018), 028
doi:10.1088/1475-7516/2018/04/028
[arXiv:1711.09478 [hep-th]].


\bibitem{Bergshoeff:2015tra}
E.~A.~Bergshoeff, D.~Z.~Freedman, R.~Kallosh and A.~Van Proeyen,
 ``Pure de Sitter Supergravity,''
Phys. Rev. D \textbf{92}, no.8, 085040 (2015)
[erratum: Phys. Rev. D \textbf{93}, no.6, 069901 (2016)]
doi:10.1103/PhysRevD.93.069901
[arXiv:1507.08264 [hep-th]].
F.~Hasegawa and Y.~Yamada,
 ``Component action of nilpotent multiplet coupled to matter in 4 dimensional $ \mathcal{N}=1 $ supergravity,''
JHEP \textbf{10}, 106 (2015)
doi:10.1007/JHEP10(2015)106
[arXiv:1507.08619 [hep-th]].

\bibitem{Moore:1998zu}
G.~W.~Moore,
``Attractors and arithmetic,''
[arXiv:hep-th/9807056 [hep-th]].
S.~Ferrara, K.~Hayakawa and A.~Marrani,
 ``Lectures on Attractors and Black Holes,''
Fortsch. Phys. \textbf{56}, no.10, 993-1046 (2008)
doi:10.1002/prop.200810569
[arXiv:0805.2498 [hep-th]].
S.~Bellucci, S.~Ferrara, R.~Kallosh and A.~Marrani,
``Extremal Black Hole and Flux Vacua Attractors,''
Lect. Notes Phys. \textbf{755}, 115-191 (2008)
[arXiv:0711.4547 [hep-th]].

\bibitem{Freedman:2012zz}
D.~Z.~Freedman and A.~Van Proeyen,
 ``Supergravity,''
Cambridge Univ. Press, 2012,
ISBN 978-1-139-36806-3, 978-0-521-19401-3
doi:10.1017/CBO9781139026833


\bibitem{Goldstein:2005hq}
K.~Goldstein, N.~Iizuka, R.~P.~Jena and S.~P.~Trivedi,
 ``Non-supersymmetric attractors,''
Phys. Rev. D \textbf{72}, 124021 (2005)
doi:10.1103/PhysRevD.72.124021
[arXiv:hep-th/0507096 [hep-th]].

\bibitem{Ceresole:2007wx}
A.~Ceresole and G.~Dall'Agata,
``Flow Equations for Non-BPS Extremal Black Holes,''
JHEP \textbf{03}, 110 (2007)
doi:10.1088/1126-6708/2007/03/110
[arXiv:hep-th/0702088 [hep-th]].

\bibitem{Boruch:2023gfn}
J.~Boruch, L.~V.~Iliesiu, S.~Murthy and G.~J.~Turiaci,
``New forms of attraction: Attractor saddles for the black hole index,''
[arXiv:2310.07763 [hep-th]].

\bibitem{Cassani:2025sim}
D.~Cassani and S.~Murthy,
``Quantum black holes: supersymmetry and exact results,''
[arXiv:2502.15360 [hep-th]].



\bibitem{Kallosh:1996uy}
R.~Kallosh and B.~Kol,
 ``E(7) symmetric area of the black hole horizon,''
Phys. Rev. D \textbf{53}, R5344-R5348 (1996)
doi:10.1103/PhysRevD.53.R5344
[arXiv:hep-th/9602014 [hep-th]].
M.~Cvetic and C.~M.~Hull,
 ``Black holes and U duality,''
Nucl. Phys. B \textbf{480}, 296-316 (1996)
doi:10.1016/S0550-3213(96)00449-X
[arXiv:hep-th/9606193 [hep-th]].
V.~Balasubramanian, F.~Larsen and R.~G.~Leigh,
 ``Branes at angles and black holes,''
Phys. Rev. D \textbf{57}, 3509-3528 (1998)
doi:10.1103/PhysRevD.57.3509
[arXiv:hep-th/9704143 [hep-th]].

\bibitem{Ferrara:1997ci}
S.~Ferrara and J.~M.~Maldacena,
 ``Branes, central charges and U duality invariant BPS conditions,''
Class. Quant. Grav. \textbf{15}, 749-758 (1998)
doi:10.1088/0264-9381/15/4/004
[arXiv:hep-th/9706097 [hep-th]].

\bibitem{Gaillard:1981rj}
M.~K.~Gaillard and B.~Zumino,
``Duality Rotations for Interacting Fields,''
Nucl. Phys. B \textbf{193}, 221-244 (1981)
doi:10.1016/0550-3213(81)90527-7

\bibitem{deWit:2002vt}
B.~de Wit, H.~Samtleben and M.~Trigiante,
``On Lagrangians and gaugings of maximal supergravities,''
Nucl. Phys. B \textbf{655}, 93-126 (2003)
doi:10.1016/S0550-3213(03)00059-2
[arXiv:hep-th/0212239 [hep-th]].
B.~de Wit, H.~Samtleben and M.~Trigiante,
``The Maximal D=4 supergravities,''
JHEP \textbf{06}, 049 (2007)
doi:10.1088/1126-6708/2007/06/049
[arXiv:0705.2101 [hep-th]].

\bibitem{Bossard:2011tq}
G.~Bossard, P.~S.~Howe, K.~S.~Stelle and P.~Vanhove,
 ``The vanishing volume of D=4 superspace,''
Class. Quant. Grav. \textbf{28}, 215005 (2011)
doi:10.1088/0264-9381/28/21/215005
[arXiv:1105.6087 [hep-th]].

\bibitem{Henneaux:2017kbx}
M.~Henneaux, B.~Julia, V.~Lekeu and A.~Ranjbar,
``A note on \textquoteleft{}gaugings\textquoteright{} in four spacetime dimensions and electric-magnetic duality,''
Class. Quant. Grav. \textbf{35}, no.3, 037001 (2018)
doi:10.1088/1361-6382/aa9fd5
[arXiv:1709.06014 [hep-th]].
C.~Hillmann,
``E(7)(7) invariant Lagrangian of d=4 N=8 supergravity,''
JHEP \textbf{04}, 010 (2010)
doi:10.1007/JHEP04(2010)010
[arXiv:0911.5225 [hep-th]].
G.~Bossard, C.~Hillmann and H.~Nicolai,
``E7(7) symmetry in perturbatively quantised N=8 supergravity,''
JHEP \textbf{12}, 052 (2010)
doi:10.1007/JHEP12(2010)052
[arXiv:1007.5472 [hep-th]].
C.~Bunster and M.~Henneaux,
``Sp(2n,R) electric-magnetic duality as off-shell symmetry of interacting electromagnetic and scalar fields,''
PoS \textbf{HRMS2010}, 028 (2010)
doi:10.22323/1.109.0028
[arXiv:1101.6064 [hep-th]].

\bibitem{Kallosh:2010kk}
R.~Kallosh,
 ``The Ultraviolet Finiteness of N=8 Supergravity,''
JHEP \textbf{12}, 009 (2010)
doi:10.1007/JHEP12(2010)009
[arXiv:1009.1135 [hep-th]].
R.~Kallosh and T.~Ortin,
 ``New E77 invariants and amplitudes,''
JHEP \textbf{09}, 137 (2012)
doi:10.1007/JHEP09(2012)137
[arXiv:1205.4437 [hep-th]].
R.~Kallosh,
 ``The Action with Manifest E7 Type Symmetry,''
JHEP \textbf{05}, 109 (2019)
doi:10.1007/JHEP05(2019)109
[arXiv:1812.08087 [hep-th]].
M.~Gunaydin and R.~Kallosh,
 ``Supersymmetry constraints on U-duality invariant deformations of $N \geq 5$ Supergravity,''
JHEP \textbf{09}, 105 (2019)
doi:10.1007/JHEP09(2019)105
[arXiv:1812.08758 [hep-th]].
R.~Kallosh and Y.~Yamada,
 ``Deformation of d = 4, $ \mathcal{N} $\ensuremath{\geq} 5 supergravities breaks nonlinear local supersymmetry,''
JHEP \textbf{06}, 156 (2023)
doi:10.1007/JHEP06(2023)156
[arXiv:2304.10514 [hep-th]].



\bibitem{Kallosh:2002gf}
R.~Kallosh, A.~D.~Linde, S.~Prokushkin and M.~Shmakova,
 ``Supergravity, dark energy and the fate of the universe,''
Phys. Rev. D \textbf{66}, 123503 (2002)
doi:10.1103/PhysRevD.66.123503
[arXiv:hep-th/0208156 [hep-th]].





\bibitem{Volkov:1973ix}
D.~V.~Volkov and V.~P.~Akulov,
 ``Is the Neutrino a Goldstone Particle?,''
Phys. Lett. B \textbf{46}, 109-110 (1973)
doi:10.1016/0370-2693(73)90490-5





\bibitem{Linde:1984ir}
A.~D.~Linde,
 ``The Inflationary Universe,''
Rept. Prog. Phys. \textbf{47}, 925-986 (1984)
doi:10.1088/0034-4885/47/8/002
A.~D.~Sakharov,
 ``Cosmological Transitions With a Change in Metric Signature,''
Sov. Phys. JETP \textbf{60}, 214-218 (1984)
doi:10.1070/PU1991v034n05ABEH002502;
S.~Weinberg,
``Anthropic Bound on the Cosmological Constant,''
Phys. Rev. Lett. \textbf{59}, 2607 (1987)
doi:10.1103/PhysRevLett.59.2607;

\bibitem{Ferrara:2013rsa}
S.~Ferrara, R.~Kallosh, A.~Linde and M.~Porrati,
``Minimal Supergravity Models of Inflation,''
Phys. Rev. D \textbf{88}, no.8, 085038 (2013)
doi:10.1103/PhysRevD.88.085038
[arXiv:1307.7696 [hep-th]].


\bibitem{Carrasco:2015rva}
J.~J.~M.~Carrasco, R.~Kallosh and A.~Linde,
 ``Cosmological Attractors and Initial Conditions for Inflation,''
Phys. Rev. D \textbf{92}, no.6, 063519 (2015)
doi:10.1103/PhysRevD.92.063519
[arXiv:1506.00936 [hep-th]].
J.~J.~M.~Carrasco, R.~Kallosh and A.~Linde,
 ``$\alpha $-Attractors: Planck, LHC and Dark Energy,''
JHEP \textbf{10}, 147 (2015)
doi:10.1007/JHEP10(2015)147
[arXiv:1506.01708 [hep-th]].




\bibitem{Cicoli:2024bxw}
M.~Cicoli, A.~Grassi, O.~Lacombe and F.~G.~Pedro,
``Chiral global embedding of Fibre Inflation with $\overline{\rm D3}$ uplift,''
[arXiv:2412.08723 [hep-th]].

\bibitem{Kallosh:2017wku}
R.~Kallosh, A.~Linde, D.~Roest, A.~Westphal and Y.~Yamada,
 ``Fibre Inflation and $\alpha$-attractors,''
JHEP \textbf{02}, 117 (2018)
doi:10.1007/JHEP02(2018)117
[arXiv:1707.05830 [hep-th]].





\bibitem{Kallosh:2015sea}
R.~Kallosh,
 ``Matter-coupled de Sitter Supergravity,''
Theor. Math. Phys. \textbf{187}, no.2, 695-705 (2016)
doi:10.1134/S0040577916050068
[arXiv:1509.02136 [hep-th]].

\bibitem{Farakos:2018sgq}
F.~Farakos, A.~Kehagias and A.~Riotto,
 ``Liberated $ \mathcal{N} $ = 1 supergravity,''
JHEP \textbf{06}, 011 (2018)
doi:10.1007/JHEP06(2018)011
[arXiv:1805.01877 [hep-th]].

\bibitem{Jang:2020cbe}
H.~Jang and M.~Porrati,
``Constraining Liberated Supergravity,''
Phys. Rev. D \textbf{103}, no.2, 025008 (2021)
doi:10.1103/PhysRevD.103.025008
[arXiv:2010.06789 [hep-th]].



\bibitem{CKLR} J.J. Carrasco, R. Kallosh, A. Linde, D. Roest,  Work in Progress




\end{thebibliography}
\end{document}